\documentclass[aps,prb,twocolumn,superscriptaddress]{revtex4-2}

\usepackage{amsmath , amssymb}
\usepackage{mathrsfs}
\usepackage[dvipdfmx]{graphicx}
\usepackage{bm}
\usepackage{braket}
\usepackage{multirow}
\usepackage{color}
\usepackage[colorlinks=true, citecolor=blue]{hyperref}

\newcommand{\del}{\partial}
\newcommand{\diff}{\mathrm{d}}
\newcommand{\imag}{\mathrm{Im}\,}

\newcommand{\real}{\mathrm{Re}\,}
\newcommand{\trace}{\mathrm{Tr}\,}
\newcommand{\imu}{\mathrm{i}}
\newcommand{\epn}{\mathrm{e}}
\newcommand{\sgn}{\mathrm{sgn}\,}
\newcommand{\ua}{\uparrow}
\newcommand{\da}{\downarrow}
\newcommand{\dg}{\dagger}
\newcommand{\la}{\langle}
\newcommand{\ra}{\rangle}
\newcommand{\al}{\alpha}
\newcommand{\be}{\beta}
\newcommand{\sg}{\sigma}
\newcommand{\om}{\omega}

\newcommand{\T}{\mathrm{T}}
\newcommand{\nt}{\notag \\}

\newcommand{\mrm}[1]{\mathrm{#1}}
\newcommand{\mcal}[1]{\mathcal{#1}}

\begin{document}

\title{
Odd-frequency pairing of Bogoliubov quasiparticles in superconductor junction
}

\author{Tatsuya Miki}
\affiliation{Department of Physics, Saitama University, Shimo-Okubo, Saitama 338-8570, Japan}
\author{Yukio Tanaka}
\affiliation{Department of Applied Physics, Nagoya University, Nagoya 464-8603, Japan}
\affiliation{Research Center for Crystalline Materials Engineering, Nagoya University, Nagoya 464-8603, Japan}
\author{Shun Tamura}
\affiliation{Faculty for Physics and Astronomy (TP4), Universit\"{a}t W\"{u}rzburg, Am Hubland, D-97074 W\"{u}rzburg, Germany}
\author{Shintaro Hoshino}
\affiliation{Department of Physics, Saitama University, Shimo-Okubo, Saitama 338-8570, Japan}

\date{\today}

\begin{abstract}

We study a superconductor Josephson junction with a Bogoliubov Fermi surface, employing McMillan's Green's function technique. 
The low-energy degrees of freedom are described by spinless fermions (bogolons), where the characteristic feature appears as an odd-frequency pair potential.
The differential equation of the Green's function is reduced to the eigenvalue problem of the non-Hermitian effective Hamiltonian.
The physical quantities such as the density of states and pair amplitude are then extracted from the obtained Green's function.
We find that the zero energy local density of states at the interface decreases as the relative phase of the Josephson junction increases.
This decrease is accompanied by the generation of an even-frequency pair amplitude near the interface.
We also clarify that the $\pi$-junction-like current phase relation is realized in terms of bogolons.
In contrast to conventional $s$-wave superconductor junctions, where even-frequency pairs dominate in the bulk and odd-frequency pairs are generated near the interface, our findings illuminate the distinct behaviors of junctions with Bogoliubov Fermi surfaces.
We further explore spatial dependences of these physical quantities systematically using quasiclassical Green's functions.

\end{abstract}

\maketitle

\section{Introduction}

The superconductors (SCs) with gapless fermionic excitations, known as Bogoliubov Fermi surfaces (BFSs) \cite{Volovik89-1, Volovik89-2, Agterberg17, Brydon18}, have been the subject of both theoretical \cite{Volovik93, Liu03, Gubankova05, Yuan18, Sumita19, Menke19, Autti20, Suh20, Setty20, Setty20-prb, Lapp20, Oh20, Tamura20, Timm21, Timm21-2, Jiang21, Miki21, Hoshino22_proc, Link20, Link20-2, Herbut20, deFarias20, Kim21, Dutta21, Oh21, Burset15, Banerjee22, Kobayashi22, Kitamura22, Cao23, Wu23, Ohashi23, Miki24} and experimental \cite{Sato18, Shibauchi20, Zhu21, Nagashima22, Mizukami23} studies.
BFSs have intriguing physics due to their potential to exhibit characteristics distinct from the Fermi surface consisting of electrons in normal metals \cite{Oh20, Tamura20, Miki21, Hoshino22_proc}.
Such an exotic state of quantum matter is potentially realized in Fe(Se,S), where the remaining density of states (DOS) and gapless quasiparticle behaviors below the transition temperature are observed \cite{Sato18, Shibauchi20, Nagashima22, Mizukami23}.
Residual DOS is observed in other superconductors \cite{Schuberth92}, which are also candidate systems with BFSs.

Our previous works have demonstrated that BFSs host purely odd-frequency Cooper pairs composed of Bogoliubov quasiparticles (bogolons), which is a distinct feature absent in conventional SCs \cite{Miki21, Miki24}.
Specifically, we focused on the bulk state of bogolons near the BFS, which are described by spinless fermions \cite{Tamura20, Miki21, Hoshino22_proc, Miki23_proc}. 
We have evaluated the self-energy of bogolons by considering the impurity and interaction effects, where a standard perturbative technique is employed as used in conventional Fermi liquid theory.
Since the number of bogolons is not a conserved quantity in contrast to that of electrons in the normal metal, the anomalous part of the self-energy, i.e., the pair potential, is finite, which makes the bogolon system different from normal Fermi liquid. 
The time dependence of this pair potential has a purely odd functional form \cite{Berezinskii74}, which is interpreted as the generation of the odd-frequency Cooper pair composed of bogolons in bulk.

Up to now, pursuing the odd-frequency pairing has been an important issue in strongly correlated systems, and there have been theoretical proposals in (multi-channel) Kondo lattice models \cite{Emery92, Coleman93, Zachar96, Jarrell96, Anders02, Hoshino14-1, Hoshino14-2, Otsuki15, Tsvelik16, Tsvelik19, Iimura19, deFarias20, Iimura21, deCarvalho21, Coleman22}, itinerant correlated electron models \cite{Bulut93, Vojta99, Fuseya03, Shigeta09, Shigeta11, Yanagi12, Fukui18, Schrodi21, Misu23}, disordered systems \cite{Kirkpatrick91, Belitz99, Santos20, Zyuzin22}, electron-phonon coupled systems \cite{Kusunose12, Matsumoto12}, SCs with BFS \cite{Kim21, Dutta21, Miki21} and other systems \cite{Balatsky92, Abrahams95, Hotta09, Iwasaki24}.
If the standard relation $F^\dg_{12}(\imu \omega_n) = F_{21}^*(-\imu\omega_n)$ for the anomalous Green's function is taken, the generation of pure odd-frequency pairs in bulk has a difficulty from the viewpoints of stability \cite{Heid95, Miki23_proc}. 
Also, the pure odd-frequency pairing with another relation $F^\dg_{12}(\imu \omega_n) = -F_{21}^*(-\imu\omega_n)$ \cite{Solenov09, Kusunose11} has a difficulty when it coexists with the odd-frequency pairing \cite{Fominov15, Matsubara21} generated by the external symmetry breaking from conventional even-frequency one \cite{Bergeret05, TanakaPRL07a, TanakaPRL07b, TanakaPRB07, Tanaka12, Matsumoto13, Tamura19, Linder19, Cayao20, Tamura21, Tanaka24}.
On the other hand, the odd-frequency pairing of bogolons discussed in this paper is more naturally induced by the self-energy effect for the systems with BFS \cite{Miki21, Miki24}, where the standard relation $F^\dg_{12}(\imu \omega_n) = F_{21}^*(-\imu\omega_n)$ holds in bulk.

Thus, the low-energy bogolon model is a suitable platform to study the physical properties of the odd-frequency Cooper pair \cite{Miki21, Miki23_proc}. 
In contrast to the previous discussions focused on bulk properties \cite{Miki21, Miki23_proc, Miki24}, it is noteworthy that the translational and inversion symmetries are broken at surfaces and interfaces. 
Namely, odd-frequency pairs can be induced at the surface or interface of conventional even-frequency SCs as a result of lack of translational symmetry \cite{TanakaPRB07,TanakaPRL07b,Tanaka12}.
Hence, it is interesting to study {\it induced even-frequency pairs} at the interface of the {\it odd-frequency} SC.

In this paper, we study the junction of the SC with the BFS based on the bogolon model. 
The schematic figure is illustrated in Fig.~\ref{fig:concept} (a).
As shown in the next section, we begin with the Gor'kov equation with different self-energies used for left- ($x < 0$) and right-side ($x > 0$) systems.
Here, we use the techniques of non-Hermitian quantum mechanics \cite{Brody14, Kawabata19} and McMillan's formalism for Green's function \cite{McMillan68}, which has been used in the conventional SC junctions with even-frequency pair potential \cite{Furusaki91, Tanaka96, Tanaka97, Kashiwaya00, Burset15, Lu18, Tanaka24}.
We study a spatial dependence of the physical quantities such as the local density of states based on the Green functions, and also quasiclassical Green's function, which extracts a slowly varying component.
Since the relation between physical quantities in terms of bogolon and experimental observables is not trivial and depends on specific details of each superconductor, we focus on properties of bogolons in this paper as a first step to understanding the junction with BFS.
We emphasize that the bogolon junction is regarded as a Josephson junction of bulk odd-frequency pairing state, which has never been explored and is a foundation to understanding superconductor junctions with BFS.

The rest part of this paper is organized as follows.
In Sec.~\ref{sec:mcmillan}, we introduce Green's function following McMillan's method.
Before showing the results for the system with the BFS, we first summarize the result of the $s$-wave SC junction for reference in Sec.~\ref{sec:swave}.
Section~\ref{sec:result} provides the result of Green's function and physical quantities at the interface of the bogolon junction.
In Sec.~\ref{sec:quasiclassical}, we evaluate the quasiclassical Green's function to study the slowly-varying spatial component of both even and odd-frequency pair amplitudes.
We summarize the paper in Sec.~\ref{sec:summary}.
The connection between bogolon and original electronic degrees of freedom is explained in Appendix~\ref{sec:app_origin}.
The detailed calculation of McMillan Green's function is given in Appendix~\ref{sec:app_mcmillan}.
The detailed results for the conventional spin-singlet $s$-wave SC case are listed in Appendix~\ref{sec:app_swave} as a reference.
The specific forms of physical quantities in quasiclassical representations are given in Appendix~\ref{sec:app_quasiclassical}.

\section{McMillan Green's function \label{sec:mcmillan}}
\subsection{Setup of bogolon junction}

\begin{figure}[tbp]
    \centering
    \includegraphics[width=8.6cm]{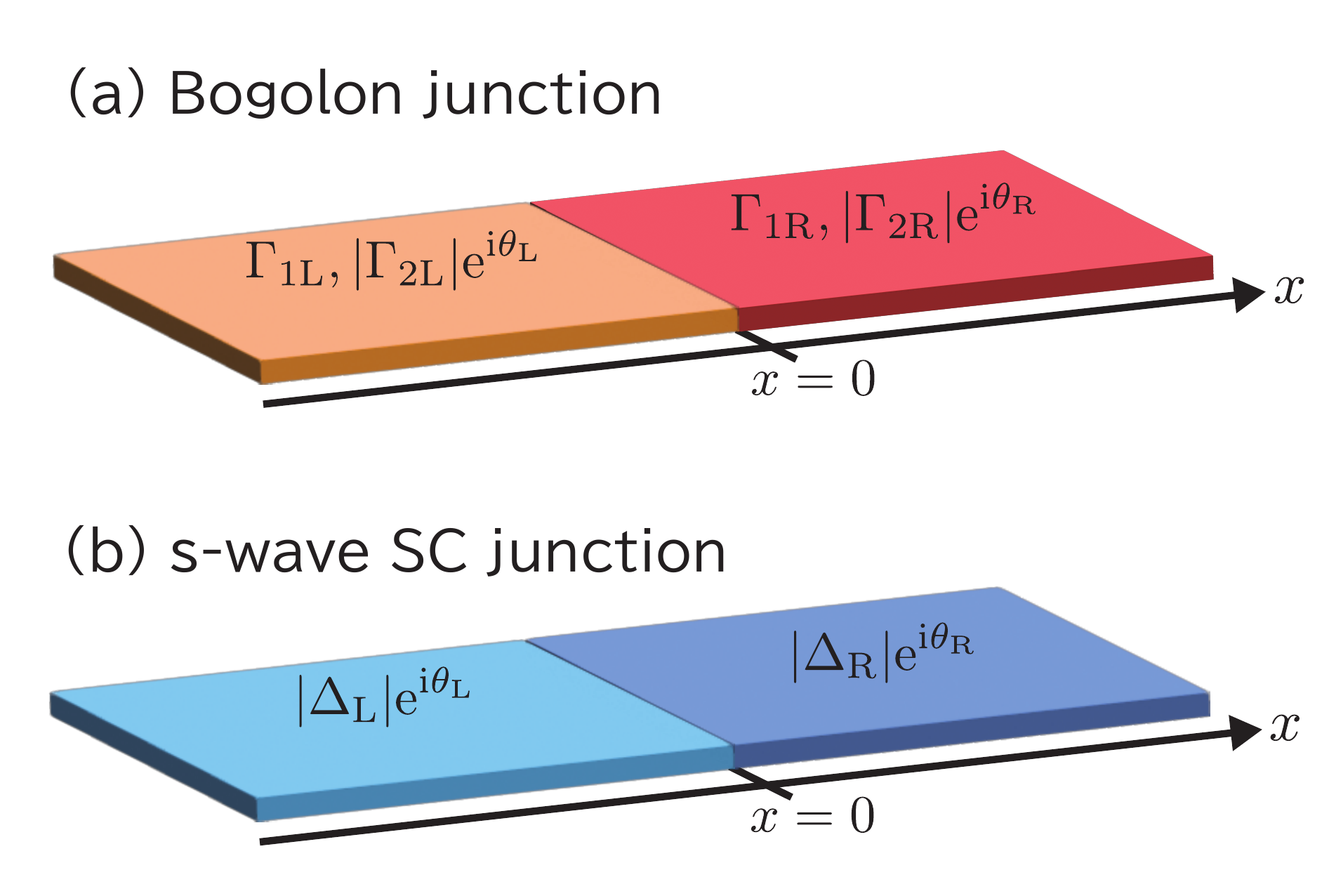}
    \caption{Schematic figure of one-dimensional SC junction. We consider two systems: (a) Bogolon junction
    and (b) $s$-wave SC junction.
    }
    \label{fig:concept}
\end{figure}

We consider a one-dimensional SC junction by combining the systems with BFSs, where low-energy behaviors are described by spinless bogolons \cite{Tamura20, Miki21, Hoshino22_proc, Miki23_proc} (bogolon junction).
We also consider the conventional $s$-wave SC junction in order to discuss the unique properties of the bogolon junction by comparing the two systems.
Figure~\ref{fig:concept} shows schematics of these two systems.
The interface is located at $x = 0$.
Although we focus on one-dimensional systems, our formulation can be extended to a three-dimensional junction system straightforwardly by considering quantum numbers $(k_y,k_z)$ with translational symmetry along $y$ and $z$ directions.
The self-energies take different values between the left side ($x < 0$) and the right side ($x > 0$).
In the following of this paper, we concentrate mainly on the bogolon model.
The results for the conventional $s$-wave SC are summarized in Sec.~\ref{sec:swave} and  Appendix~\ref{sec:app_swave}.

We consider the BFS in a time-reversal symmetry broken superconductor \cite{Agterberg17, Brydon18, Setty20, Setty20-prb}, where elementary excitations near the BFS are described by spinless fermions \cite{Tamura20, Miki21, Hoshino22_proc, Miki23_proc}.
Since the pair of bogolons is realized by taking into account the anomalous self-energy \cite{Miki21, Miki24}, the Green's function formalism is suitable for the description of bogolon junctions.
The Green's function is defined in terms of bogolons' creation/annihilation operators as
\begin{align}
    \hat G(\tau, x,x')  = - \la \mathcal T \vec \al(\tau, x) \vec \al^\dg (x') \ra,
\end{align}
where $\mcal T$ is a time ordering, and $A(\tau)$ is a imaginary-time Heisenberg representation of an operator $A$.
The vector operator is given by $\vec \al(x) = (\al(x), \al^\dg(x))^\T$ where $\al(x)$ is an annihilation operator for the low-energy bogolon.
The hat symbols ($\, \hat{}\, $) indicate $2\times 2$ matrices in Nambu space.
We require the Green's function to satisfy the two (left- and right-) Gor'kov equations in the fermionic Matsubara frequency ($\omega_n$) domain, which are explicitly given by
\begin{align}
    &[\imu\om_n\hat 1 - \hat H_0(x) - \hat \Sigma(\imu\om_n, x)] \hat G(\imu\om_n, x, x') = \delta(x - x') \label{eq:gorkov_1}, \\
    &\hat G(\imu\om_n, x, x') [\imu\om_n\hat 1 - \hat H_0(x') - \hat \Sigma(\imu\om_n, x')] = \delta(x - x') \label{eq:gorkov_2},
\end{align}
where $\hat \Sigma(\imu\om_n, x)$ is a self-energy, and $\hat 1$ is a two dimensional identity matrix.
$\hat H_0(x)$ in Eqs.~\eqref{eq:gorkov_1} and \eqref{eq:gorkov_2} is a Hamiltonian of ideal bogolon gas with the barrier potential:
\begin{align}
    \hat H_0(x) = \Big[-\frac{\hbar^2}{2m}\frac{\diff^2}{\diff x^2} - \mu + V_{\mrm B}\delta(x) \Big]\hat \tau^z,
\end{align}
where $m$ and $\mu$ are effective mass and chemical potential of bogolons, respectively.
In the above equation, we introduce the barrier potential $V_{\mrm B}$ at the interface $x = 0$.
Since we consider the same $\hat H_0(x)$ between the right side ($x > 0$) and the left side ($x < 0$), we assume that both sides are the same materials.

The model of bogolon for bulk was introduced in our previous works \cite{Miki21,Miki24}.
The low-energy bogolons form pure odd-frequency pairs in bulk.
Here, we phenomenologically introduce the spatial dependence of the self-energy $\hat \Sigma(\imu\om_n, x)$ given by
\begin{align}
    \hat \Sigma(\imu\om_n, x) = 
    -\imu
    \begin{pmatrix}
        \Gamma_1(x) & \Gamma_2(x) \\
        \Gamma_2(x)^\ast & \Gamma_1(x)
    \end{pmatrix} \sgn\om_n.
    \label{eq:self_ene_general}
\end{align}
The presence of the sign function represents the odd-frequency pair potential in the off-diagonal part.
The spatial dependence of $\Gamma_1(x)$ and $\Gamma_2(x)$ are given by
\begin{align}
    &\Gamma_1(x) = 
    \begin{cases}
        \Gamma_{1\mrm{L}} \qquad (x < 0), \\
        \Gamma_{1\mrm{R}} \qquad (x > 0),
    \end{cases} \label{eq:gamma_1} \\
    &\Gamma_2(x) = 
    \begin{cases}
        \Gamma_{2\mrm{L}} = |\Gamma_{2\mrm{L}}| \epn^{\imu\theta_{\mrm L}} \qquad (x < 0), \\
        \Gamma_{2\mrm{R}} = |\Gamma_{2\mrm{R}}| \epn^{\imu\theta_{\mrm R}} \qquad (x > 0),
    \end{cases} \label{eq:gamma_2}
\end{align}
where $\Gamma_{1 r}$ ($r = \mrm R, \mrm L$) is a positive constant and corresponds to the quasiparticle dumping of bogolon.
On the other hand, $\Gamma_{2 r}$ is a complex number.
We note that $\Gamma_{1 r}$ and $\Gamma_{2 r}$ must satisfy the relation $\Gamma_{1 r} > |\Gamma_{2 r}|$ ($0 \leq |\Gamma_{2r}| / \Gamma_{1r} < 1$) to guarantee the positive DOS in bulk \cite{Miki21, Hoshino22_proc, Miki23_proc}. 
The schematic figure of our setup is illustrated in Fig.~\ref{fig:concept} (a).

\subsection{Origins of phase of pair potential}

Let us comment on the correspondence of our bogolon model with the original electron degrees of freedom, on which bogolons are based.
While the origins of the pair potential of bogolon have already been discussed in Refs.~\cite{Miki21, Hoshino22_proc, Miki24}, we briefly revisit the discussion with a particular focus on the origin of the phase of the pair potential for bogolons, which is crucial in the Josephson junction.
Let us consider the impurity scattering as an origin of the self-energy.
The phase of pair potential for bogolon can be understood by the following two steps.
The first step is to recognize that impurity effects on anomalous self-energy for bogolons enter through $\al^\dg \al$-type diagonal scattering ($U_1$) and $\al^\dg \al^\dg$-type off-diagonal one ($U_2$) \cite{Miki24}.
The second step is to recognize that the latter scattering potential is composed of the product of two types of wave functions $u$ (electron wave function) and $v$ (hole wave function).
Therefore, the superconducting phase of the original electron's pair potential ($\arg uv$) is reflected in $U_2$, and is then also inherited to bogolon's pair potential $\Gamma_2$.
On the other hand, $U_1$ is composed of products of $u^* u$ and $v^* v$, both of which do not carry the phase of superconducting pair potential.
The more detailed expressions are given in Appendix~\ref{sec:app_origin} and Ref.~\cite{Miki21}.

\subsection{McMillan Green's function \label{sec:mcmillan_formalism}}

In order to solve the Gor'kov equations given in Eqs.~\eqref{eq:gorkov_1} and \eqref{eq:gorkov_2}, we employ McMillan formalism \cite{McMillan68, Furusaki91, Tanaka96, Tanaka97, Kashiwaya00, Burset15, Lu18, Tanaka24}.
We consider the following two eigenequations for a given $\om_n$:
\begin{align}
    &[\hat H_0(x) + \hat \Sigma(\imu\om_n, x)] \Psi(x) = \imu\om_n \Psi(x), \label{eq:wfc_1} \\
    &\tilde \Psi(x)^\T [\hat H_0(x) + \hat \Sigma(\imu\om_n, x)] = \imu\om_n \tilde \Psi(x)^\T. \label{eq:wfc_2}
\end{align}
Once these eigenequations are solved, the Green's function is then expressed as the bilinear form of the eigenfunctions $\Psi$ and $\tilde \Psi$ as will be shown later [see Eq.~\eqref{eq:green_func_mcmillan}].
In Eqs.~\eqref{eq:wfc_1} and \eqref{eq:wfc_2}, $\hat H_0(x) + \hat\Sigma(\imu\om_n, x)$ can be regarded as an effective Hamiltonian which is not necessarily Hermitian.
Then, as we will discuss in Sec.~\ref{sec:solution}, we evaluate the Green's function using a biorthogonal basis employed for non-Hermitian Hamiltonian systems \cite{Brody14, Kawabata19}.
Note that the junctions with the non-Hermitian system have been discussed in previous studies \cite{Kornich22, Kornich23}.
On the other hand, the non-Hermiticity discussed in this paper is derived from self-energy.
Therefore, the relation $\hat \Sigma^\ast(-\imu\om_n, x) = \hat \Sigma^\dg(\imu\om_n, x)$ derived from standard Lehmann representation is always satisfied in our formulation, from which any unphysical results, such as imaginary physical quantities, are excluded.
In this paper, the non-Hermitian formalism is just a methodology to solve Eqs.~\eqref{eq:wfc_1} and \eqref{eq:wfc_2}.

In this subsection, we consider the case of $\om_n > 0$.
Since the self-energies does not depend on the spatial coordinate $x$ once $x>0$ or $x<0$ is specified [see Eqs.~\eqref{eq:gamma_1} and \eqref{eq:gamma_2}], the eigenfunction for Eq.~\eqref{eq:wfc_1} can be expressed by the plane wave, as in an ordinary scattering problem.
Although the following argument in this subsection is essentially the same as the case for the conventional SC junction, whose $\hat \Sigma$ given by Eq.~\eqref{eq:delta_mat} is Hermitian \cite{McMillan68, Furusaki91, Tanaka96, Tanaka97, Kashiwaya00, Burset15, Lu18, Tanaka24}, it can also be applied to our bogolon model with Eq.~\eqref{eq:self_ene_general}.

\begin{figure}[tbp]
    \centering
    \includegraphics[width=8.6cm]{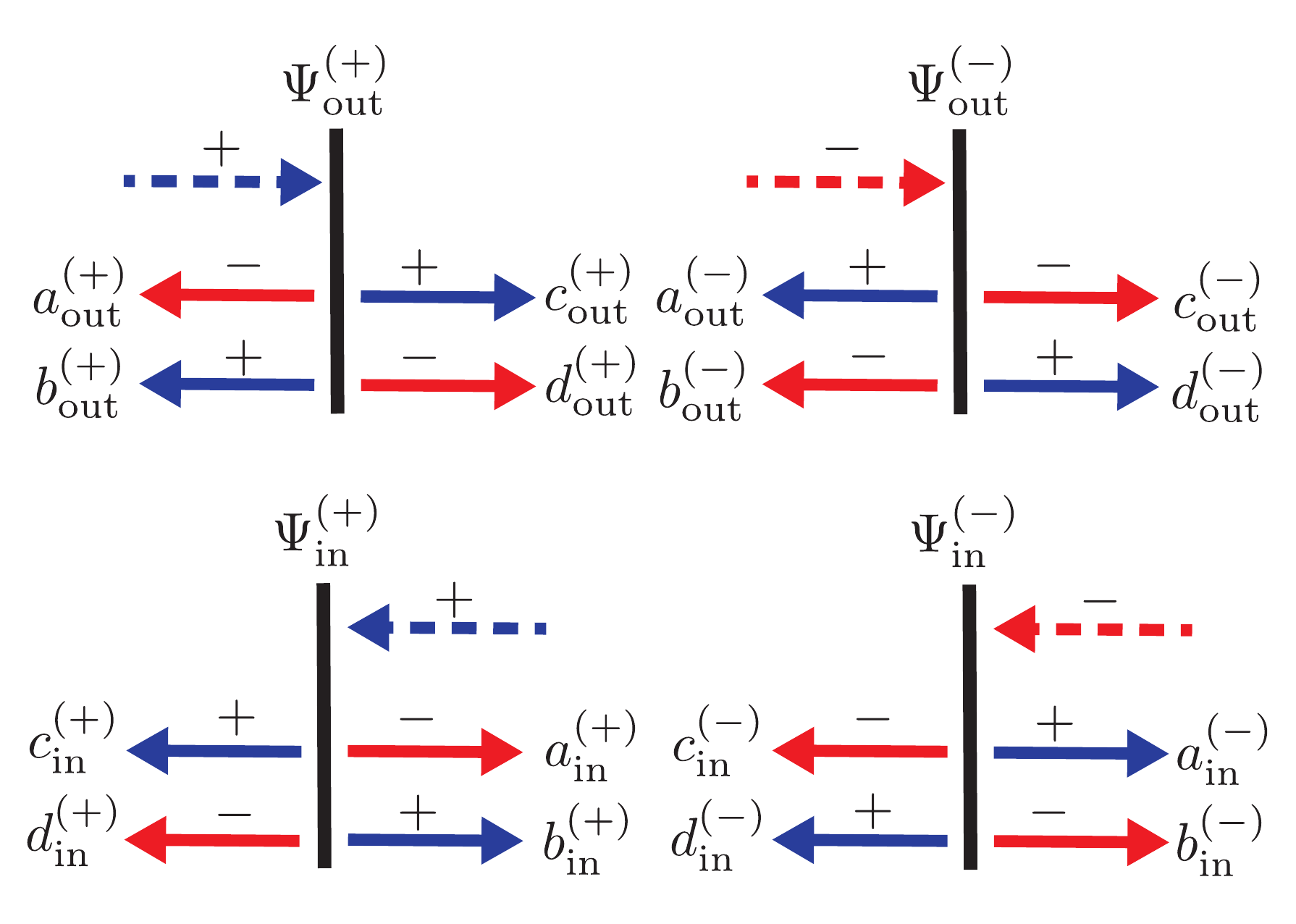}
    \caption{Four types of eigenfunctions.
    The red ($+$) arrows and blue ($-$) arrows indicate the particle- and the hole-like plane waves, respectively.
    The dashed lines represent the incident waves.
    The directions of arrows indicate the group velocity.
    The sign in the superscript of $\Psi$, $a$, $b$, $c$, and $d$ indicates particle- $(+)$ or hole-like particle $(-)$ incident.
    }
    \label{fig:wfc}
\end{figure}

We need to consider the four types of independent eigenfunctions [outgoing/incoming (out/in), particle/hole ($+/-$)] shown in Fig.~\ref{fig:wfc} \cite{Furusaki91, Tanaka96, Tanaka97, Kashiwaya00, Burset15, Lu18, Tanaka24}, which depend on the incident plane wave indicated by dashed lines.
For example, the outgoing particle wave function $\Psi_{\mrm{out}}^{(+)}$ is given by
\begin{align}
    \Psi_{\mrm{out}}^{(+)}(x) = 
    \begin{cases}
        \epn^{\imu k_{\mrm L}^+ x} 
        \begin{pmatrix}
            u_{\mrm L}^+ \\
            v_{\mrm L}^+
        \end{pmatrix}
        + a_{\mrm{out}}^{(+)} \epn^{\imu k_{\mrm L}^- x}
        \begin{pmatrix}
            u_{\mrm L}^- \\
            v_{\mrm L}^-
        \end{pmatrix} \\
        \qquad + b_{\mrm{out}}^{(+)} \epn^{-\imu k_{\mrm L}^+ x}
        \begin{pmatrix}
            u_{\mrm L}^+ \\
            v_{\mrm L}^+
        \end{pmatrix}  \qquad (x < 0), \\
        c_{\mrm{out}}^{(+)} \epn^{\imu k_{\mrm R}^+ x}
        \begin{pmatrix}
            u_{\mrm R}^+ \\
            v_{\mrm R}^+
        \end{pmatrix}
        + d_{\mrm{out}}^{(+)} \epn^{-\imu k_{\mrm R}^- x}
        \begin{pmatrix}
            u_{\mrm R}^- \\
            v_{\mrm R}^-
        \end{pmatrix}  \quad (x > 0),
    \end{cases} \label{eq:wfc}
\end{align}
which is a linear combination of bulk solutions.
In the above expression, the wave number $k_r^\pm$, kinetic energy $\Omega_r(\imu\om_n)$, and bulk wave functions $u_r^\pm, v_r^\pm$ ($r = \mrm{L},  \mrm{R}$) satisfy the following relations:
\begin{align}
    &k_r^\pm = \sqrt{(2m/\hbar^2)[\mu \pm \Omega_r(\imu\om_n)]}, \\
    &\Big[\pm\Omega_r(\imu\om_n) \hat \tau^z + \hat \Sigma_r(\imu\om_n)\Big]
    \begin{pmatrix}
        u_r^\pm \\
        v_r^\pm
    \end{pmatrix}
    = \imu\om_n
    \begin{pmatrix}
        u_r^\pm \\
        v_r^\pm
    \end{pmatrix}. \label{eq:eigen_bulk}
\end{align}
The solution of Eq.~\eqref{eq:eigen_bulk} is discussed in the next subsection in detail.
Note that $\Omega_r(\imu\om_n)$ can also be represented in another form as shown later [see Eq.~\eqref{eq:omega}].
In Eq.~\eqref{eq:eigen_bulk}, $(u_r^{+(-)}, v_r^{+(-)})^{\mathrm{T}}$ corresponds to the particle (hole) eigenfunction. 
The eigenfunctions corresponding to $\tilde \Psi$ in Eq.~\eqref{eq:wfc_2} are denoted as $(\tilde u_r^\pm, \tilde v_r^\pm)^{\mathrm{T}}$.

As seen in Fig.~\ref{fig:wfc}, the coefficients $a_{\mrm{out}}^{(\pm)}$ and $a_{\mrm{in}}^{(\pm)}$ correspond to the reflection from particle to hole.
Then, we call this type of reflection as {\it Andreev reflection} of bogolons, which is analogous to the conventional SC junctions.
Similarly, we call ``$b_{\mrm{out}}^{(\pm)}, b_{\mrm{in}}^{(\pm)}$'' and ``$c_{\mrm{out}}^{(\pm)}, c_{\mrm{in}}^{(\pm)}$'' in Eq.~\eqref{eq:wfc} and Fig.~\ref{fig:wfc} as normal reflection and normal transmission, respectively. 
The transmission from particle to hole is reflected in ``$d_{\mrm{out}}^{(\pm)}, d_{\mrm{in}}^{(\pm)}$''.
The specific expressions of $a_{\mrm{out}}^{(\pm)}, a_{\mrm{in}}^{(\pm)}, b_{\mrm{out}}^{(\pm)}, b_{\mrm{in}}^{(\pm)}, c_{\mrm{out}}^{(\pm)}, c_{\mrm{in}}^{(\pm)}, d_{\mrm{out}}^{(\pm)}$, and $ d_{\mrm{in}}^{(\pm)}$ are determined by the boundary conditions at $x = 0$ \cite{McMillan68, Furusaki91, Tanaka96, Tanaka97, Kashiwaya00, Burset15, Lu18, Tanaka24}, which are given by
\begin{align}
    &\Psi_{\mrm{out}}^{(\pm)}(x = +0^+) = \Psi_{\mrm{out}}^{(\pm)}(x = - 0^+), \label{eq:wfc_cond_1} \\
    &\left.\frac{\diff \Psi_{\mrm{out}}^{(\pm)}(x)}{\diff x}\right|_{x = 0^+} - \left.\frac{\diff \Psi_{\mrm{out}}^{(\pm)}(x)}{\diff x}\right|_{x = -0^+} 
    = \left. \frac{2mV_{\mrm B}}{\hbar^2} \Psi_{\mrm{out}}^{(\pm)}(x) \right|_{x = 0}.
    \label{eq:wfc_cond_2}
\end{align}
We also impose the same conditions on $\Psi_{\mrm{in}}^{(\pm)}$.
These conditions are similar to those in the scattering problem with the barrier potential of quantum mechanics.
The specific forms of $a_{\mrm{out}}^{(\pm)}$, $a_{\mrm{in}}^{(\pm)}$, $b_{\mrm{out}}^{(\pm)}$, $b_{\mrm{in}}^{(\pm)}$, $c_{\mrm{out}}^{(\pm)}$, $c_{\mrm{in}}^{(\pm)}$, $d_{\mrm{out}}^{(\pm)}$, and $d_{\mrm{in}}^{(\pm)}$ are listed in Eqs.~\eqref{eq:aoutz}--\eqref{eq:dinz}.
We also evaluate the eigenfunction $\tilde \Psi_{\mrm{out}}(x), \tilde \Psi_{\mrm{in}}(x)$ in Eq.\eqref{eq:wfc_2} in a manner similar to $\Psi_{\mrm{out}}(x), \Psi_{\mrm{in}}(x)$.

Using the four types of eigenfunctions described in Fig.~\ref{fig:wfc}, the Green's function is expressed in the following form \cite{Furusaki91, Tanaka96, Tanaka97, Kashiwaya00, Burset15, Lu18, Tanaka24}:
\begin{widetext}
\begin{align}
    \hat G(\imu\om_n, x, x') = 
    \begin{cases}
        \al_1 \Psi_{\mrm{in}}^{(+)}(x)\tilde{\Psi}_{\mrm{out}}^{(+)}(x')^\T 
        + \al_2 \Psi_{\mrm{in}}^{(+)}(x)\tilde{\Psi}_{\mrm{out}}^{(-)}(x')^\T 
        + \al_3 \Psi_{\mrm{in}}^{(-)}(x)\tilde{\Psi}_{\mrm{out}}^{(+)}(x')^\T 
        + \al_4 \Psi_{\mrm{in}}^{(-)}(x)\tilde{\Psi}_{\mrm{out}}^{(-)}(x')^\T  \qquad (x < x'), \\
        \be_1 \Psi_{\mrm{out}}^{(+)}(x)\tilde{\Psi}_{\mrm{in}}^{(+)}(x')^\T 
        + \be_2 \Psi_{\mrm{out}}^{(-)}(x)\tilde{\Psi}_{\mrm{in}}^{(+)}(x')^\T 
        + \be_3 \Psi_{\mrm{out}}^{(+)}(x)\tilde{\Psi}_{\mrm{in}}^{(-)}(x')^\T 
        + \be_4 \Psi_{\mrm{out}}^{(-)}(x)\tilde{\Psi}_{\mrm{in}}^{(-)}(x')^\T \qquad (x > x'),
    \end{cases} \label{eq:green_func_mcmillan}
\end{align}
\end{widetext}
where coefficients $\al_1, \cdots, \al_4$ and $\be_1, \cdots, \be_4$ are determined by the boundary conditions at $x = 0$, which are given by
\begin{align}
    \hat G(\imu\om_n, x \to x' + 0^+, x') = \hat G(\imu\om_n, x \to x' + 0^-, x') \label{eq:boundary_3}
\end{align}
and
\begin{align}
\hspace{-5mm}
    \left.\frac{\del}{\del x}\hat G(\imu\om_n, x, x')\right|_{x=x'+0^+} - \left.\frac{\del}{\del x}\hat G(\imu\om_n, x, x')\right|_{x=x'-0^+} = \frac{2m}{\hbar^2} \hat \tau^z. \label{eq:boundary_4}
\end{align}
We note that Eq.~\eqref{eq:boundary_4} is derived from the Gor'kov equations in Eqs.~\eqref{eq:gorkov_1} and \eqref{eq:gorkov_2}.
The Green's function for $\om_n < 0$ can be evaluated by the conjugate relation
\begin{align}
    \hat G(-\imu\om_n, x, x') = \hat G^\dg(\imu\om_n, x', x), \label{eq:hermite}
\end{align}
which is derived from the Lehmann representation of the Green's function.

We emphasize again that Eq.~\eqref{eq:green_func_mcmillan} can be used for both bogolon (non-Hermite case) and $s$-wave SC (Hermite case).
In the next subsection, we evaluate the specific form of $u_r^\pm$, $v_r^\pm$, $\tilde u_r^\pm$, and $\tilde v_r^\pm$ by using a biorthogonal basis, which can be applied to both the Hermite ($s$-wave SC junction) and the non-Hermite cases (boglon junction).

\subsection{Solution of auxiliary non-Hermitian problem \label{sec:solution}}

Now we consider the concrete eigenvalue problem, in which the characteristic feature of bogolons, i.e. Non-Hermitian nature of effective Hamiltonian, is reflected.
The following discussion can be used for both $\om_n > 0$ and $\om_n < 0$.
First we write down Eq.~\eqref{eq:eigen_bulk} for the $r$-side ($r = \mrm R$ for $x > 0$ and $r = \mrm L$ for $x < 0$) as 
\begin{align}
    \begin{pmatrix}
        \pm\Omega_r(\imu\om_n) - \imu\Gamma_{1r}\sgn\om_n & S_r(\imu\om_n) \\
        S_r^+(\imu\om_n) & \mp\Omega_r(\imu\om_n) - \imu\Gamma_{1r}\sgn\om_n
    \end{pmatrix} 
    \begin{pmatrix}
        u_r^\pm \\
        v_r^\pm
    \end{pmatrix} \nt
    = \imu\om_n
    \begin{pmatrix}
        u_r^\pm \\
        v_r^\pm
    \end{pmatrix}.
    \label{eq:ham_sig_1}
\end{align}
For the bogolon junction, we set $\Gamma_{1r} > 0, S_r(\imu\om_n) = -\imu\Gamma_{2r}\sgn\om_n, S_r^+(\imu\om_n) = -\imu\Gamma_{2r}^\ast\sgn\om_n$ 
[If we set $\Gamma_{1r} = 0$ and $S_r(\imu\om_n) = S_r^+(\imu\om_n)^\ast = \Delta_r$, Eq.~\eqref{eq:ham_sig_1} applies to the $s$-wave SC.].
Since the matrix in Eq.~\eqref{eq:ham_sig_1} is non-Hermitian, it is necessary to consider the following Hermite conjugate version:
\begin{align}
    \begin{pmatrix}
        \pm\Omega_r(\imu\om_n)^\ast + \imu\Gamma_{1r}\sgn\om_n & S_r^+(\imu\om_n)^\ast \\
        S_r(\imu\om_n)^\ast & \mp\Omega_r(\imu\om_n)^\ast + \imu\Gamma_{1r}\sgn\om_n
    \end{pmatrix} 
    \begin{pmatrix}
        \tilde u_r^{\pm\ast} \\
        \tilde v_r^{\pm\ast}
    \end{pmatrix} \nt
    = (\imu\om_n)^\ast
    \begin{pmatrix}
        \tilde u_r^{\pm\ast} \\
        \tilde v_r^{\pm\ast}
    \end{pmatrix}.
    \label{eq:ham_sig_2}
\end{align}
The wave functions $u_r^\pm$, $\tilde u_r^\pm$, $v_r^\pm$, and $\tilde v_r^\pm$ satisfy the following biorthogonal condition
\footnote{
Note that, for the Hermite case with $s$-wave SC, $u_r^\pm$ and $v_r^\pm$ are identical to $\tilde u_r^{\pm\ast}$ and $\tilde v_r^{\pm\ast}$, respectively, but it is not the case for bogolon model.
}
, which is used in the context of non-Hermitian quantum mechanics \cite{Brody14, Kawabata19}:
\begin{align}
    \tilde u_r^\pm u_r^\pm + \tilde v_r^\pm v_r^\pm = 1. \label{eq:biorthogonal}
\end{align}
From Eqs.~\eqref{eq:ham_sig_1} and \eqref{eq:ham_sig_2}, $\Omega_r(\imu\om_n)$ is determined as
\begin{align}
    \Omega_r(\imu\om_n) = \sqrt{(\imu\om_n + \imu\Gamma_{1r}\sgn\om_n)^2 - S_r(\imu\om_n) S_r^+(\imu\om_n)}. \label{eq:omega}
\end{align}
We note that $\Omega_r(\imu\om_n)$ is an even function of $\om_n$ in the bogolon model (this is also true for the $s$-wave SC).

As discussed in Sec.~\ref{sec:mcmillan_formalism}, the coefficients $\al_1, \cdots, \al_4$ and $\be_1, \cdots, \be_4$ in Eq.~\eqref{eq:green_func_mcmillan} are determined by the boundary conditions for the Green's function in Eqs.~\eqref{eq:boundary_3} and \eqref{eq:boundary_4}.
The specific form of $\al_1, \cdots, \al_4$ and $\be_1, \cdots, \be_4$ are listed in Eqs.~\eqref{eq:al1}--\eqref{eq:be4}.
Using these coefficients, the Green's function Eq.~\eqref{eq:green_func_mcmillan} for $x, x' < 0$ is reduced to
\begin{widetext}
\begin{align}
    \hat G(\imu\om_n, x, x') 
    &= \frac{m (\imu\om_n + \imu\Gamma_{1\mrm{L}}\sgn\om_n)}{\imu k_{\mrm F} \hbar^2 \Omega_{\mrm{L}}(\imu\om_n)} 
    \Bigg[ \Big(\epn^{\imu k_{\mrm L}^+ |x - x'|} + \tilde b_{\mrm{out}}^{(+)} \epn^{-\imu k_{\mrm L}^+ (x + x')} \Big)
    \begin{pmatrix}
        u_{\mrm L}^+ \tilde u_{\mrm L}^+ & u_{\mrm L}^+ \tilde v_{\mrm L}^+ \\
        v_{\mrm L}^+ \tilde u_{\mrm L}^+ & v_{\mrm L}^+ \tilde v_{\mrm L}^+
    \end{pmatrix}
    + \tilde a_{\mrm{out}}^{(+)} \epn^{-\imu k_{\mrm L}^+ x + \imu k_{\mrm L}^- x'}
    \begin{pmatrix}
        u_{\mrm L}^+ \tilde u_{\mrm L}^- & u_{\mrm L}^+ \tilde v_{\mrm L}^- \\
        v_{\mrm L}^+ \tilde u_{\mrm L}^- & v_{\mrm L}^+ \tilde v_{\mrm L}^-
    \end{pmatrix} \nt
    &+ \Big(\epn^{-\imu k_{\mrm L}^- |x - x'|} + \tilde b_{\mrm{out}}^{(-)} \epn^{\imu k_{\mrm L}^- (x + x')} \Big)
    \begin{pmatrix}
        u_{\mrm L}^- \tilde u_{\mrm L}^- & u_{\mrm L}^- \tilde v_{\mrm L}^- \\
        v_{\mrm L}^- \tilde u_{\mrm L}^- & v_{\mrm L}^- \tilde v_{\mrm L}^-
    \end{pmatrix}
    + \tilde a_{\mrm{out}}^{(-)} \epn^{\imu k_{\mrm L}^- x - \imu k_{\mrm L}^+ x'}
    \begin{pmatrix}
        u_{\mrm L}^- \tilde u_{\mrm L}^+ & u_{\mrm L}^- \tilde v_{\mrm L}^+ \\
        v_{\mrm L}^- \tilde u_{\mrm L}^+ & v_{\mrm L}^- \tilde v_{\mrm L}^+
    \end{pmatrix}
    \Bigg], \label{eq:green_func_mcmillan_general}
\end{align}
where $k_{\mrm F}$ is a Fermi wavevector of bogolon and $\tilde{a}_{\mrm{out}}^{(\pm)}$ and $\tilde{b}_{\mrm{out}}^{(\pm)}$ are coefficients in $\tilde{\Psi}_{\mrm{out}}^{(\pm)}(x)$.
This expression, which includes six terms, provides a clearer physical meaning than Eq.~\eqref{eq:green_func_mcmillan} \cite{Furusaki91}.
Namely, the first line describes particle behavior, while the second line describes hole behavior.
The contributions from the bulk are represented by the first and fourth terms. 
The second and the fifth terms, which are proportional to $\tilde b_{\mrm{out}}^{(\pm)}$, correspond to the normal reflection.
The third and sixth terms proportional to $\tilde a_{\mrm{out}}^{(\pm)}$ represent the contributions from the Andreev reflection.

\subsection{Concrete form of Green's function \label{sec:gfunc_full}}

In the following of this paper, we focus on the bogolon junction with $\Gamma_{1\mrm L} = \Gamma_{1\mrm R}$ and $|\Gamma_{2\mrm L}| = |\Gamma_{2\mrm R}|$.
The concrete form of the Green's function in this subsection is one of the central results of this paper.
Using the functional forms of $u_{\mrm L}^\pm$, $v_{\mrm L}^\pm$, $\tilde u_{\mrm L}^{\pm\ast}$, and $\tilde v_{\mrm L}^{\pm\ast}$ given in Appendix~\ref{sec:app_mcmillan}, the Green's function for $x, x' < 0$ is obtained as follows:
\begin{align}
    \hat G(\imu\om_n, x, x') 
    &= \frac{m}{2\imu k_{\mrm F} \hbar^2 \Omega_{\mrm L}(\imu\om_n)}
    \Bigg[ 
    \Big(\epn^{\imu k_{\mrm L}^+ |x - x'|} + \tilde b_{\mrm{out}}^{(+)} \epn^{-\imu k_{\mrm L}^+ (x + x')} \Big)
    \begin{pmatrix}
        \imu\om_n + \imu\Gamma_{1\mrm L}\sgn\om_n + \Omega_{\mrm L}(\imu\om_n) & -\imu \Gamma_{2\mrm L}\sgn\om_n \\
        -\imu \Gamma_{2\mrm L}^\ast\sgn\om_n & \imu\om_n + \imu\Gamma_{1\mrm L}\sgn\om_n - \Omega_{\mrm L}(\imu\om_n)
    \end{pmatrix} 
    \nt
    &+ \tilde a_{\mrm{out}}^{(+)} \epn^{-\imu k_{\mrm L}^+ x + \imu k_{\mrm L}^- x'} 
    \begin{pmatrix}
        \imu|\Gamma_{2\mrm L}|\sgn\om_n & -\epn^{\imu\theta_{\mrm L}} [\imu\om_n + \imu\Gamma_{1\mrm L}\sgn\om_n + \Omega_{\mrm L}(\imu\om_n)] \\
        -\epn^{-\imu\theta_{\mrm L}} [\imu\om_n + \imu\Gamma_{1\mrm L}\sgn\om_n - \Omega_{\mrm L}(\imu\om_n)] & \imu |\Gamma_{2\mrm L}|\sgn\om_n
    \end{pmatrix} 
    \nt
    &+ \Big(\epn^{-\imu k_{\mrm L}^- |x - x'|} + \tilde b_{\mrm{out}}^{(-)} \epn^{\imu k_{\mrm L}^- (x + x')} \Big)
    \begin{pmatrix}
        \imu\om_n + \imu\Gamma_{1\mrm L}\sgn\om_n - \Omega_{\mrm L}(\imu\om_n) & -\imu\Gamma_{2\mrm L}\sgn\om_n \\
        -\imu\Gamma_{2\mrm L}^\ast\sgn\om_n & \imu\om_n + \imu\Gamma_{1\mrm L}\sgn\om_n + \Omega_{\mrm L}(\imu\om_n)
    \end{pmatrix} \nt
    &+ \tilde a_{\mrm{out}}^{(-)} \epn^{\imu k_{\mrm L}^- x - \imu k_{\mrm L}^+ x'} 
    \begin{pmatrix}
        \imu |\Gamma_{2\mrm L}|\sgn\om_n & -\epn^{\imu\theta_{\mrm L}} [\imu\om_n + \imu\Gamma_{1\mrm L}\sgn\om_n - \Omega_{\mrm L}(\imu\om_n)] \\
        -\epn^{-\imu\theta_{\mrm L}} [\imu\om_n + \imu\Gamma_{1\mrm L}\sgn\om_n + \Omega_{\mrm L}(\imu\om_n)] & \imu |\Gamma_{2\mrm L}|\sgn\om_n
    \end{pmatrix}
    \Bigg] \label{eq:mcmillan_bog}
\end{align}
with
\begin{align}
    &\tilde a_{\mrm{out}}^{(\pm)}(\imu\om_n) 
    = \frac{-\imu|\Gamma_{2\mrm L}|\sgn\om_n[(\imu\om_n + \imu\Gamma_{1\mrm L}\sgn\om_n) \sin^2(\theta/2) \pm \imu \Omega_{\mrm L}(\imu\om_n) \sin(\theta/2)\cos(\theta/2)] }{Z^2\Omega_{\mrm L}(\imu\om_n)^2 + (\imu\om_n + \imu\Gamma_{1\mrm L} \sgn\om_n)^2 + |\Gamma_{2\mrm L}|^2 \cos^2(\theta/2)}, \label{eq:a} \\
    &\tilde b_{\mrm{out}}^{(\pm)}(\imu\om_n)
    = -\frac{Z(Z \pm \imu\sgn\om_n) \Omega_{\mrm L}(\imu\om_n)^2 }{Z^2 \Omega_{\mrm L}(\imu\om_n)^2 + (\imu\om_n + \imu\Gamma_{1\mrm L}\sgn\om_n)^2 + |\Gamma_{2\mrm L}|^2 \cos^2(\theta/2)}. \label{eq:b}
\end{align}
\end{widetext}
We have introduced the relative phase of the pair potential $\theta = \theta_{\mrm R} - \theta_{\mrm L}$ with $-\pi<\theta\leq\pi$ and defined
\begin{align}
    &Z = \frac{m V_{\mrm B}}{k_{\mrm F} \hbar^2}, \label{eq:z}
\end{align}
which represents a magnitude of delta-function barrier potential at $x=0$.

\subsection{Physical quantities \label{sec:quantities}}

\subsubsection{$s$-wave component}

Let us consider the physical quantities derived from McMillan Green's function.
From the local Green's function ($x = x'$), we define the $s$-wave component (i.e., symmetric in terms of the exchange of $x$ and $x'$) of the pair amplitude as
\begin{align}
    F_{s}(\imu\om_n, x) = \Big[\hat G(\imu\om_n, x, x)\Big]_{12}. \label{eq:fs_def}
\end{align}
We employ the Matsubara frequency representation for pair amplitudes, which is useful to recognize the even- and odd-frequency components.

We can also evaluate the local density of states (LDOS) of bogolons from the local Green's function.
The LDOS of bogolons is defined by using the diagonal components of the retarded Green's function, which is obtained through analytical continuation with respect to frequency:
\begin{align}
    D(\om, x) 
    &= -\frac{1}{\pi}\imag\trace \hat G(\om + \imu 0^+, x, x), \label{eq:ldos_def}
\end{align}
with which the probability density of bogolons is identified.

\subsubsection{$p$-wave component}

Next, we define $p$-wave component of Green's function (i.e., antisymmetric in terms of the exchange of $x$ and $x'$) as
\begin{align}
    &\hat G_{p}(\imu\om_n, x) \nt
    &= \lim_{\Delta x \to 0} \frac{1}{\Delta x} \Big[ \hat G(\imu\om_n, x + \Delta x, x) - \hat G(\imu\om_n, x, x + \Delta x) \Big] \nt
    &= \bigg( \frac{\del}{\del x} - \frac{\del}{\del x'} \bigg) \hat G(\imu\om_n, x, x') \bigg|_{x' \to x}.
    \label{eq:gp_def}
\end{align}
Inserting Eq.~\eqref{eq:mcmillan_bog} into Eq.~\eqref{eq:gp_def} and assuming $k^+ \simeq k^- \simeq k_{\mrm F}$, one finds that the bulk parts in Eq.~\eqref{eq:mcmillan_bog} vanishes.
This fact implies that $\hat G_{p}(\imu\om_n, x)$ is induced by the presence of the interface.

The $p$-wave component of the pair amplitude $F_{p}(\imu\om_n, x)$ is defined by \cite{TanakaPRB07,Tanaka12}
\begin{align}
    F_{p}(\imu\om_n, x) = \Big[\hat G_{p}(\imu\om_n, x)\Big]_{12}. \label{eq:fp_def}
\end{align}
We note that, from the Fermi-Dirac statistics, $F_s(\imu\omega_n,x)$ is an odd (even) function of $\om_n$ for bogolons ($s$-wave SC), and $F_p(\imu\omega_n,x)$ is an even (odd) function of $\om_n$ for bogolons ($s$-wave SC).

It is also worthwhile to explore the diagonal components in Nambu space of $\hat{G}_p(\imu\om_n,x)$.
The trace of $\hat{G}_p(\imu\om_n,x)$ can be regarded as quasiparticle contribution of current \cite{Furusaki91}, which is given by
\begin{align}
    J(x) 
    &= \imu\bigg\la \al^\dg(x) \frac{\del}{\del x} \al(x) - \bigg(\frac{\del}{\del x} \al^\dg(x)\bigg) \al(x) \bigg\ra \nt
    &= \frac{1}{\be} \sum_n j(\imu\om_n, x) \label{eq:j_def}
\end{align}
with
\begin{align}
    j(\imu\om_n, x) = \frac{\imu}{n_{\mrm d}} \trace \hat G_{p}(\imu\om_n, x), \label{eq:jom_def}
\end{align}
and $n_{\mrm d}$ is a factor correcting double counting: $n_{\mrm d} = 2$ for bogolon and $n_{\mrm d} = 1$ for $s$-wave SC.

\section{Summary of s-wave superconductor junction \label{sec:swave}}

Before showing the results for the bogolon junction,
here we summarize the result of the $s$-wave SC junction for reference.
The detailed setup and expressions are listed in Appendix~\ref{sec:app_swave}.
In the following of this section, we assume $|\Delta_{\mrm L}| = |\Delta_{\mrm R}|$ corresponding to the discussion in Sec.~\ref{sec:gfunc_full}.
For all figures in this section, we choose $|\Delta_{\mrm L}|/\mu = 0.01$ with $\mu = \hbar^2 k_{\mrm F}^2/2m$ and set $\mu = 1$ and $k_{\mrm F}=1$.
The specific form of the Green's function is shown in Eqs.~\eqref{eq:mcmillan_s}--\eqref{eq:b_s}.
We note that the results in this section are not completely original.

\subsection{Semi-infinite superconductor}

First, we discuss the result for the semi-infinite SC ($x < 0$), where the edge is located at $x = 0$. 
We take the limit $Z \to \infty$ in Eqs.~\eqref{eq:mcmillan_s}--\eqref{eq:b_s}.
The pair amplitude $[\hat G(\imu\om_n, x, x')]_{12}$ is given by
\begin{align}
    &[\hat G(\imu\om_n, x, x')]_{12}
    = \frac{m \Delta_{\mrm L}}{2\imu k_{\mrm F} \hbar^2 \sqrt{(\imu\om_n)^2 - |\Delta_{\mrm L}|^2}} \nt
    &\times\Big(\epn^{\imu k_{\mrm L}^+ |x - x'|} - \epn^{-\imu k_{\mrm L}^+ (x + x')} + \epn^{-\imu k_{\mrm L}^- |x - x'|} - \epn^{\imu k_{\mrm L}^- (x + x')} \Big). \label{eq:gswave_sf}
\end{align}
$[\hat G(\imu\om_n, x, x')]_{12}$ is the even function of frequency, which is the same frequency dependence as the bulk.
This frequency dependence implies the absence of Andreev reflection in Eq.~\eqref{eq:gswave_sf}.
In this way, $s$-wave pair potential, which does not have spatial dependence, does not induce the $p$-wave component at the edge \cite{TextTanaka2021}. 
Although odd-frequency pairs are not induced by the constant pair potential, spatially varying pair potential, $\Delta(x)\neq \mrm{Const.}$, induces the odd-frequency $p$-wave pair at the surface~\cite{TanakaPRB07,Tanaka12}.

\subsection{Superconductor junction without barrier potential}

\begin{figure}[tbp]
    \centering
    \includegraphics[width=8.6cm]{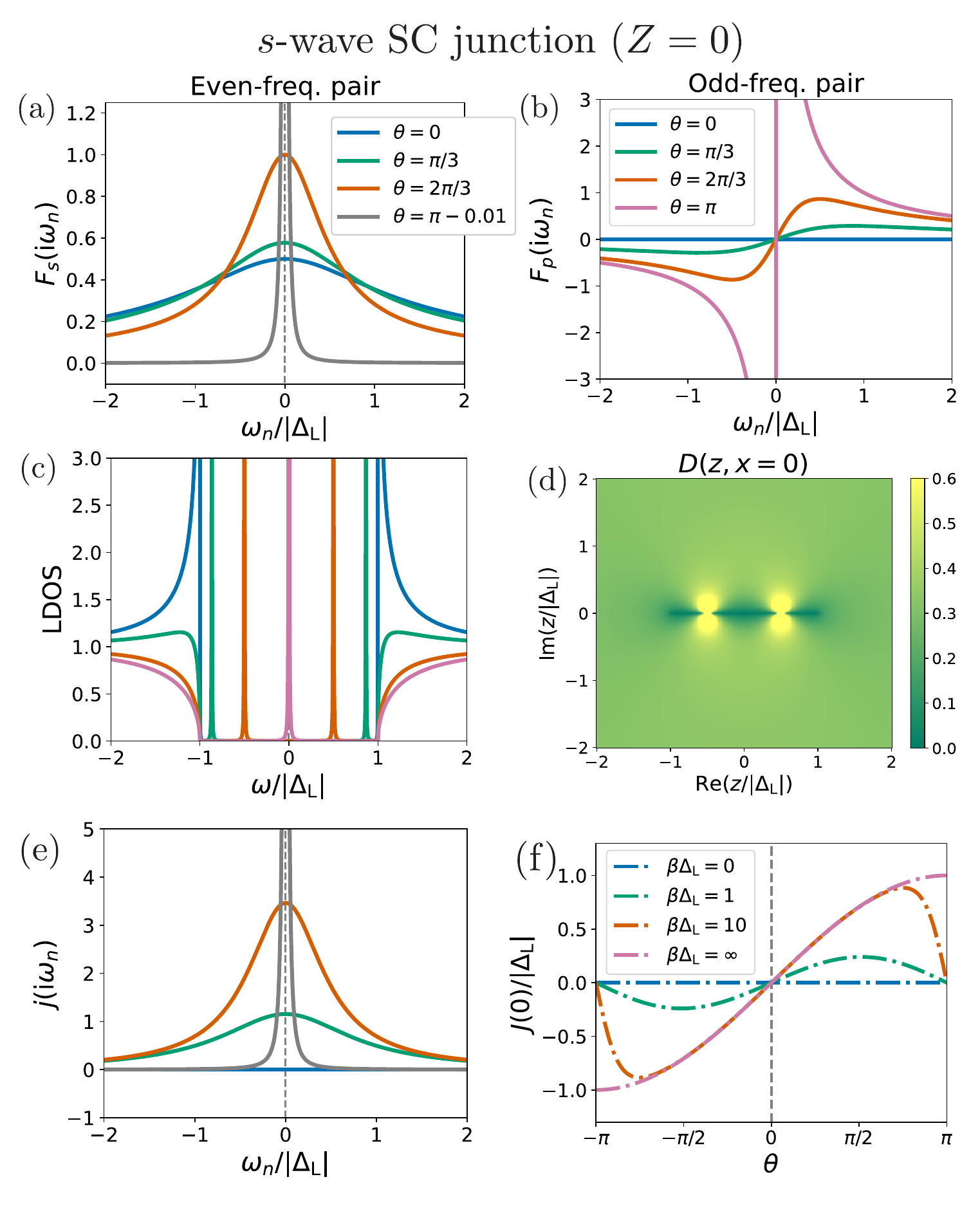}
    \caption{
    Physical quantities at $x = 0$ for $s$-wave SC junction without barrier potential.
    (a) The $s$-wave component of the pair amplitude,
    (b) the $p$-wave component of the pair amplitude, 
    (c) the LDOS normalized by its value in normal state ($\Delta_{\mrm L} = 0$), 
    (d) the LDOS in the complex plane for $\theta = 2\pi/3$, which is defined by the extension of $\om$ to the complex energy plane $z$, i.e., $D(\om, x=0) \to D(z, x=0)$, 
    (e) $j(\imu\omega_{n})=j(\imu\omega_{n},x=0)$,
    and (f) the Josephson current $J(x=0)$.
    Figure legend of (c) is the same as that of (a), and the legend of (e) is the same as that of (b).
    }
    \label{fig:mcmillan_s1}
\end{figure}

Next, we show the result of the SC junction without the barrier potential, i.e., $Z = 0$.
We assume $x, x' < 0$ in the following of this section.
The expressions of the physical quantities are listed below.
The $s$-wave component of pair amplitude defined by Eq.~\eqref{eq:fs_def} is given by
\begin{align}
    &F_s(\imu\om_n, x) 
    = \frac{\Delta_{\mrm L} m}{\imu k_{\mrm F} \hbar^2 \Omega_{\mrm L}(\imu\om_n)}
    \Bigg[ 1
    - \epn^{-\imu (k_{\mrm L}^+ - k_{\mrm L}^-) x} \nt
    &\times\frac{(\imu\om_n)^2\sin^2(\theta/2) + \imu\Omega_{\mrm L}(\imu\om_n)^2 \sin(\theta/2) \cos(\theta/2)}{(\imu\om_n)^2 - |\Delta_{\mrm L}|^2 \cos^2(\theta/2)} 
    \Bigg]. \label{eq:localf_s}
\end{align}
The numerical result at $x = 0$ is shown in Fig.~\ref{fig:mcmillan_s1} (a).
We note that the phase of $F_s(\imu\om_n, x = 0)$ is independent of $\om_n$. 
For the plot, we choose the phase such that $F_s(\imu\om_n, x = 0)$ becomes real [this choice also applies to Figs.~\ref{fig:mcmillan_s1} (b), \ref{fig:mcmillan_s2} (a), and \ref{fig:mcmillan_s2} (b)].
The functional form is substantially modified from bulk as the relative phase $\theta =\theta_{\mrm R} - \theta_{\mrm L}$ is increased.
The peak of the pair amplitude becomes sharper with increasing $\theta$ and shows delta-function-like behavior at $\theta = \pi - \delta$ with $\delta \ll 1$.
The specific form of the pair amplitude at $x = 0$ is given by
\begin{align}
    F_s(\imu\om_n, x=0) 
    &= \frac{\imu \Delta_{\mrm L} m \sqrt{\om_n^2 + |\Delta_{\mrm L}|^2}}{k_{\mrm F} |\Delta_{\mrm L}| \hbar^2} \frac{\delta|\Delta_{\mrm L}|/2}{\om_n^2 + \delta^2 |\Delta_{\mrm L}|^2/4} 
\end{align}
for $\theta = \pi - \delta$ with $\delta \ll 1$.
The right-hand side shows the presence of the Lorentzian with the width $\delta |\Delta_{\rm L}|/2$.

The $p$-wave component of pair amplitude defined by Eq.~\eqref{eq:fp_def} becomes finite due to translational symmetry breaking \cite{TanakaPRB07, Tanaka12}:
\begin{align}
    F_p(\imu\om_n, x)
    &= \frac{2\imu\Delta_{\mrm L} m}{\hbar^2} 
    \frac{ \imu\om_n \sin(\theta/2) \epn^{-\imu\theta/2}}{(\imu\om_n)^2 - |\Delta_{\mrm L}|^2 \cos^2(\theta/2)} 
    \epn^{-\imu (k_{\mrm L}^+ - k_{\mrm L}^-) x}.
    \label{eq:fp_s}
\end{align}
$F_p(\imu\om_n,x=0)$ is shown in Fig.~\ref{fig:mcmillan_s1} (b).
For $\theta = \pi$, the pair amplitude diverges at $\om_n \to 0$, because the denominator in Eq.~\eqref{eq:fp_s} becomes zero at $\theta \to \pi, \om_n \to 0$.

Let us also discuss diagonal components of the Green's function.
The LDOS is given by
\begin{align}
    &D(\om, x) 
    = D_{\mrm{bulk}}(\om) \nt 
    &-\frac{2m}{k_{\mrm F} \hbar^2 \pi}\real\bigg[ \frac{\om + \imu0^+}{\Omega_{\mrm{ret}}(\omega)} 
    \frac{|\Delta_{\mrm L}|^2 \sin^2(\theta/2) \epn^{-\imu(k_{\mrm L}^+ - k_{\mrm L}^-)x}}{(\om + \imu0^+)^2 - |\Delta_{\mrm L}|^2 \cos^2(\theta/2)} \bigg]
\end{align}
with $\Omega_{\mrm{ret}}(\omega) = \Omega_{\mrm L}(\om + \imu 0^+) \sgn\om$.
We choose the branch cut of square root function in $\Omega$ along the negative side of real axis.
We have introduced the bulk DOS by
\begin{align}
    D_{\mrm{bulk}}(\om) = \frac{2m}{k_{\mrm F}\hbar^2 \pi}\real \frac{\om + \imu0^+}{\Omega_{\mrm{ret}}(\omega)}. \label{eq:bulkdos_s}
\end{align}
The LDOS of the $s$-wave SC shown in Fig.~\ref{fig:mcmillan_s1} (c) has divergence peaks, which correspond to the Andreev bound states \cite{deGennes63, TextTanaka2021}.

In Sec.~\ref{sec:wobarrier},
we will discuss a generalized LDOS by changing $\omega \in \mathbb R$ $\to$ $z \in \mathbb C$:
$D(\omega, x) \to D(z, x)$, which will allow us to understand the relation between LDOSs of $s$-wave SC and bogolon junctions.
Figure~\ref{fig:mcmillan_s1} (d) shows the LDOS in the complex plane.
The positions of the peaks originated from Andreev bound states are given by $\om_{\mrm{ABS}}(\theta) = \pm |\Delta_{\mrm L}| \sqrt{1 - \sin^2(\theta/2)}$, which appear on the real axis.
The comparison with the bogolon junction will be discussed in Sec.~\ref{sec:wobarrier}.

We turn to the discussion of $J(x)$ defined by Eq.~\eqref{eq:j_def}.
We start the discussion with $j(\imu\om_n, x)$, which shows a contribution at the frequency $\om_n$ [see Eq.~\eqref{eq:jom_def}]:
\begin{align}
    &j(\imu\om_n, x) 
    = -\frac{4m|\Delta_{\mrm L}|^2}{\hbar^2} \frac{\sin(\theta/2) \cos(\theta/2) \epn^{-\imu (k_{\mrm L}^+ - k_{\mrm L}^-) x}}{(\imu\om_n)^2 - |\Delta_{\mrm L}|^2 \cos^2(\theta/2)}. \label{eq:jom_s}
\end{align}
The numerical result is shown in Fig.~\ref{fig:mcmillan_s1} (e) for the $x=0$ case.
The peak of $j(\imu\om_n, x=0)$ becomes sharper as the relative phase $\theta$ increases and delta-function-like behavior at $\theta = \pi - \delta$ with $\delta \ll 1$, which is similar to the $s$-wave component of pair amplitude.
Especially at $x=0$, we can perform the summation of $\om_n$ in Eq.~\eqref{eq:j_def} and obtain \cite{Kulik78}
\begin{align}
    &J(x = 0) = \frac{2m|\Delta_{\mrm L}|}{\hbar^2} \sin\frac{\theta}{2} \tanh\bigg(\frac{\be|\Delta_{\mrm L}|}{2} \cos\frac{\theta}{2} \bigg) \label{eq:j_s}.
\end{align}
For zero-temperature limit $\be \to \infty$ with $\theta \neq \pi$, we obtain the simpler form:
\begin{align}
    &J(x = 0) = \frac{2m|\Delta_{\mrm L}|}{\hbar^2} \sin\frac{\theta}{2}. \label{eq:j_s_t0}
\end{align}
The current-phase relation is plotted in Fig.~\ref{fig:mcmillan_s1} (f) for $-\pi < \theta < \pi$.

Actually, Eq.~\eqref{eq:j_s} is related to the Josephson (conserving) current \cite{Kulik78}.
According to Ref.~\cite{Furusaki91}, the conserving current $I$ is evaluated as $I = J(x=0)$ for $s$-wave SC junction case, which is derived from Heisenberg equation.

\subsection{Effect of barrier potential}

\begin{figure}[tbp]
    \centering
    \includegraphics[width=8.6cm]{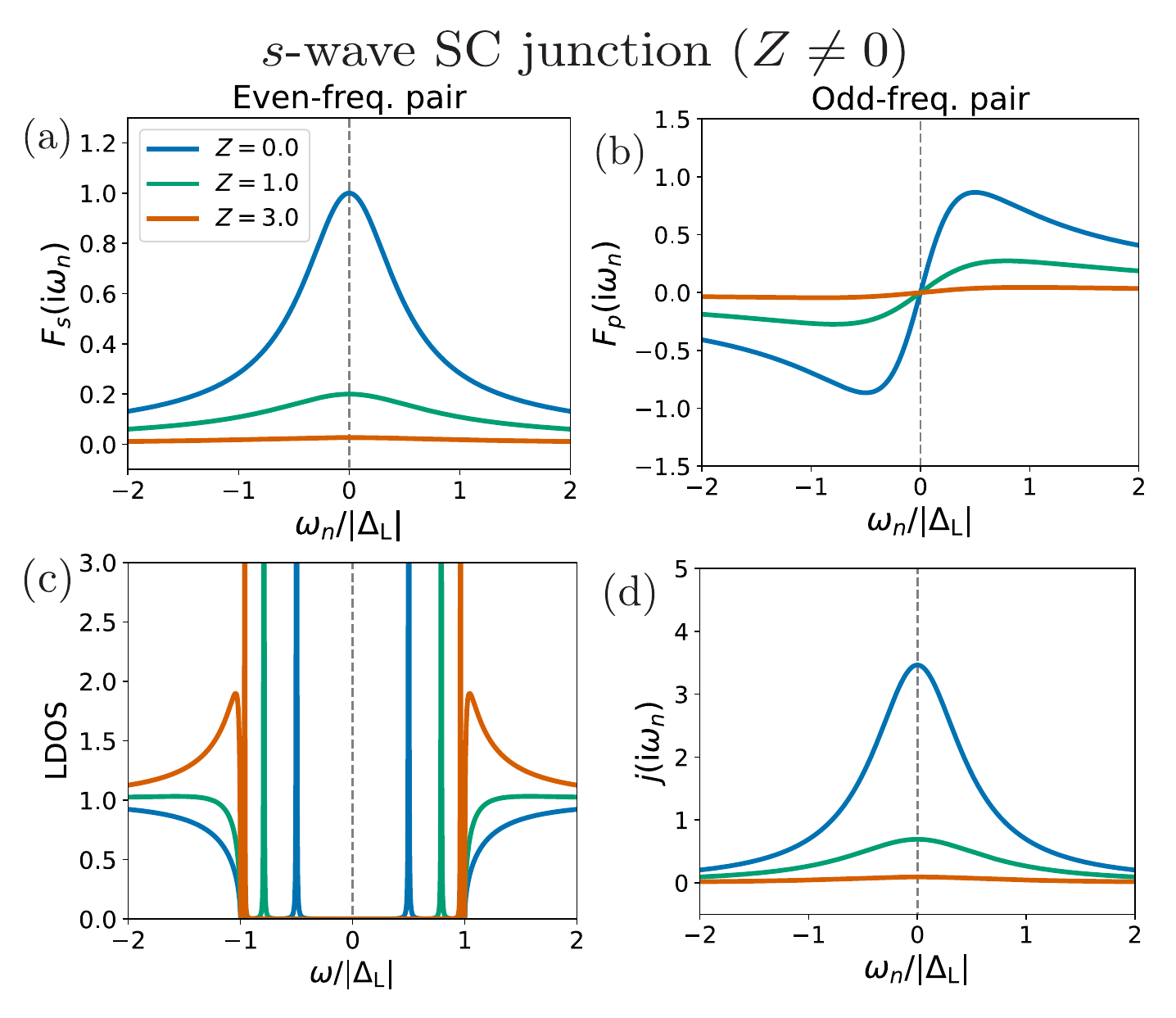}
    \caption{
    Frequency-dependence of physical quantities of $s$-wave SC at $x = 0$  for several values of $Z$.
    (a) The $s$-wave component of the pair amplitude (even-frequency), 
    (b) the $p$-wave component of the pair amplitude 
    (odd-frequency),
    (c) the LDOS normalized by its value in normal state ($\Delta_{\mrm L} = 0$), 
    and (d) $j(\imu\omega_{n})=j(\imu\omega_{n},x=0)$.
    The relative phase is chosen as $\theta = 2\pi/3$.
    }
    \label{fig:mcmillan_s2}
\end{figure}

We consider the effect of the barrier potential controlled by the parameter $Z$ defined in Eq.~\eqref{eq:z}.
Figures~\ref{fig:mcmillan_s2} (a) and (b) show the $Z$ dependence of the $s$-wave and $p$-wave components of pair amplitudes at $x = 0$, respectively.
The relative phase is chosen as $\theta = 2\pi/3$.
The delta-function-like behavior in $F_s(\imu\om_n,x=0)$ near $\theta = \pi$ can be seen only at $Z = 0$ as shown in the inset of (a).

Below, we discuss the characteristic features of the LDOS and $J(x)$ at $x = 0$ in detail.
The LDOS at $x = 0$ is expressed as follows:
\begin{align}
    &D(\om, x = 0) \nt
    &= \frac{2m}{\pi k_{\mrm F} \hbar^2} 
    \real \frac{(\om + \imu0^+) \Omega_{\mrm{ret}}(\omega)}{Z^2 \Omega_{\mrm{ret}}(\omega)^2 + (\om + \imu0^+)^2 - |\Delta|^2 \cos^2(\theta/2)}.
\end{align}
Figure~\ref{fig:mcmillan_s2} (c) shows the LDOS for $\theta = 2\pi/3$.
The position of the peak for $\omega>0$ in the LDOS moves to the higher energy as $Z$ increases.
The positions of the peaks are given by $\om_{\mrm{AB}}(\theta) = \pm |\Delta_{\mrm L}| \sqrt{1 - \sin^2(\theta/2)/(Z^2 + 1)}$
\cite{Arnold1985,Furusaki91}. 
Taking the limit $Z \to \infty$, we obtain
\begin{align}
    D(\om, x = 0) 
    &= \frac{2m}{\pi Z^2 k_{\mrm F}\hbar^2} \real \frac{\om + \imu0^+}{\Omega_{\mrm{ret}}(\omega)} \nt
    &= \frac{1}{Z^2} D_{\mrm{bulk}}(\om), \label{eq:ldos_s_zinf}
\end{align}
where $D_{\mrm{bulk}}(\om)$ is the bulk DOS defined in Eq.~\eqref{eq:bulkdos_s}.
It is notable that Eq.~\eqref{eq:ldos_s_zinf} is independent of $\theta$ and proportional to the bulk DOS in this limit.
The magnitude of the LDOS becomes smaller by the factor $1/Z^2$.

Next, we discuss the result of $J(x=0)$.
Before taking the summation of $\om_n$ in Eq.~\eqref{eq:j_def}, $j(\imu\om_n)$ is suppressed as $Z$ increases, which is shown in Fig.~\ref{fig:mcmillan_s2} (d).
We perform the summation of $\om_n$ and the expression at zero temperature limit $\be \to \infty$ is obtained as \cite{TextTanaka2021}:
\begin{align}
    J(x=0) &= \frac{m|\Delta|}{\hbar^2} \frac{1}{Z^2 + 1}\sqrt{\frac{Z^2 + 1}{Z^2 + \cos^2(\theta/2)}} \sin\theta.
\end{align}
We can confirm that the above expression reduces to Eq.~\eqref{eq:j_s_t0} in $Z \to 0$ limit.
On the other hand, for $Z \to \infty$ limit, we obtain
\begin{align}
    J(x=0) &= \frac{m|\Delta|}{\hbar^2 Z^2} \sin\theta,
\end{align}
whose $\theta$-dependence is determined by the factor $\sin\theta$ \cite{Ambegaokar63}.
We note that $J(x=0)$ is accompanied by the factor $1/Z^2$, which is the same feature as that of LDOS in the strong barrier limit.

\section{Result for bogolon junction \label{sec:result}}

With the knowledge of the conventional SC junction explained in Sec.~\ref{sec:swave}, we are now ready to discuss the bogolon junction.
In the following subsections, we apply the Green's function in Eq.~\eqref{eq:mcmillan_bog} to the specific cases.
Firstly, in the next subsection (Sec.~\ref{sec:semi_infinite}) we will provide the result of the semi-infinite system as the simplest case.
This case corresponds to the $Z\to \infty$ limit of Eq.~\eqref{eq:mcmillan_bog}.
Secondly, in Sec.~\ref{sec:wobarrier}, we will discuss the result of the bogolon junction without the barrier potential, i.e. $Z = 0$, to explore the physics of Andreev reflection of bogolon.
Finally in Sec.~\ref{sec:barrier}, we will consider the bogolon junction for $Z \neq 0$ and clarify the effect of the barrier potential.
The comparison of bogolon junction with $s$-wave SC junction is summarized in Table~\ref{tab:comparison}.
For all figures in this section, we choose $\Gamma_{1\mrm L}/\mu = 0.01$ with $\mu = \hbar^2 k_{\mrm F}^2/2m$ and set $\mu = 1$ and $k_{\mrm F}=1$.

We mainly discuss the physical quantities at the interface ($x = 0$) in this section.
The $x$-dependence can be seen more clearly using the quasiclassical Green's function, which will be discussed in the next section (Sec.~\ref{sec:quasiclassical}).

\subsection{Semi-infinite superconductor with Bogoliubov Fermi surface \label{sec:semi_infinite}}

In this subsection, we consider the semi-infinite SC ($x<0$) as the simplest non-uniform system, where the edge is located at $x=0$.
We take the limit $Z\to \infty$ in Eqs.~\eqref{eq:mcmillan_bog}--\eqref{eq:b}
\footnote{
The same expression of Green's function for the simi-infinite superconductor can be obtained considering only the normal reflection and normal transmission parts of the wave functions in Fig.~\ref{fig:wfc}.
Namely, we set $a_{\mrm{out}}^{(\pm)} = 0, b_{\mrm{out}}^{(\pm)} = 1$ in $\Psi_{\mrm{out}}^{(\pm)}$ and $c_{\mrm{in}}^{(\pm)} = 1, d_{\mrm{in}}^{(\pm)} = 0$ in $\Psi_{\mrm{in}}^{(\pm)}$, and impose the boundary condition for the wave function $\Psi_{\mrm{out}}^{(\pm)}(x = 0) = 0$.}.
The pair amplitude $[\hat G(\imu\om_n, x, x')]_{12}$ is given in the following form:
\begin{align}
    &[\hat G(\imu\om_n, x, x')]_{12}
    = \frac{-\imu\Gamma_{2\mrm L} \sgn\om_n m}{2\imu k_{\mrm F}\hbar^2 \sqrt{(\imu\om_n + \imu\Gamma_{1\mrm L}\sgn\om_n)^2 + |\Gamma_{2\mrm L}|^2}} \nt
    &\times\Big(\epn^{\imu k_{\mrm L}^+ |x - x'|} - \epn^{-\imu k_{\mrm L}^+ (x + x')} + \epn^{-\imu k_{\mrm L}^- |x - x'|} - \epn^{\imu k_{\mrm L}^- (x + x')} \Big). \label{eq:gbog_sf}
\end{align}
Since there are no Andreev reflections, the pair amplitude, i.e. $[\hat G(\imu\om_n, x, x')]_{12}$, exhibits a pure odd-frequency pairing in Eq.~\eqref{eq:gbog_sf}, which is the same frequency dependence as the bulk.
Namely, the $p$-wave component is not induced at the edge in the case of the bogolon junction without spatial dependence in the self-energy.

In order to highlight distinctive properties arising from the translational symmetry breaking at interfaces, the following subsections focus on the SC junctions and studies the physics of Andreev reflection of bogolon.

\subsection{Bogolon junction without barrier potential \label{sec:wobarrier}}

In this subsection, we consider the bogolon junction with $Z=0$.
From Eq.~\eqref{eq:b}, we have the relation $b_{\mrm{out}}^{(\pm)} = 0$, which implies that there are no normal reflections and transmissions.
Then, we can focus on the contribution of the Andereev reflection to the physical properties of the bogolon junction.

\subsubsection{Off-diagonal quantities}
\begin{figure}[tbp]
    \centering
    \includegraphics[width=8.6cm]{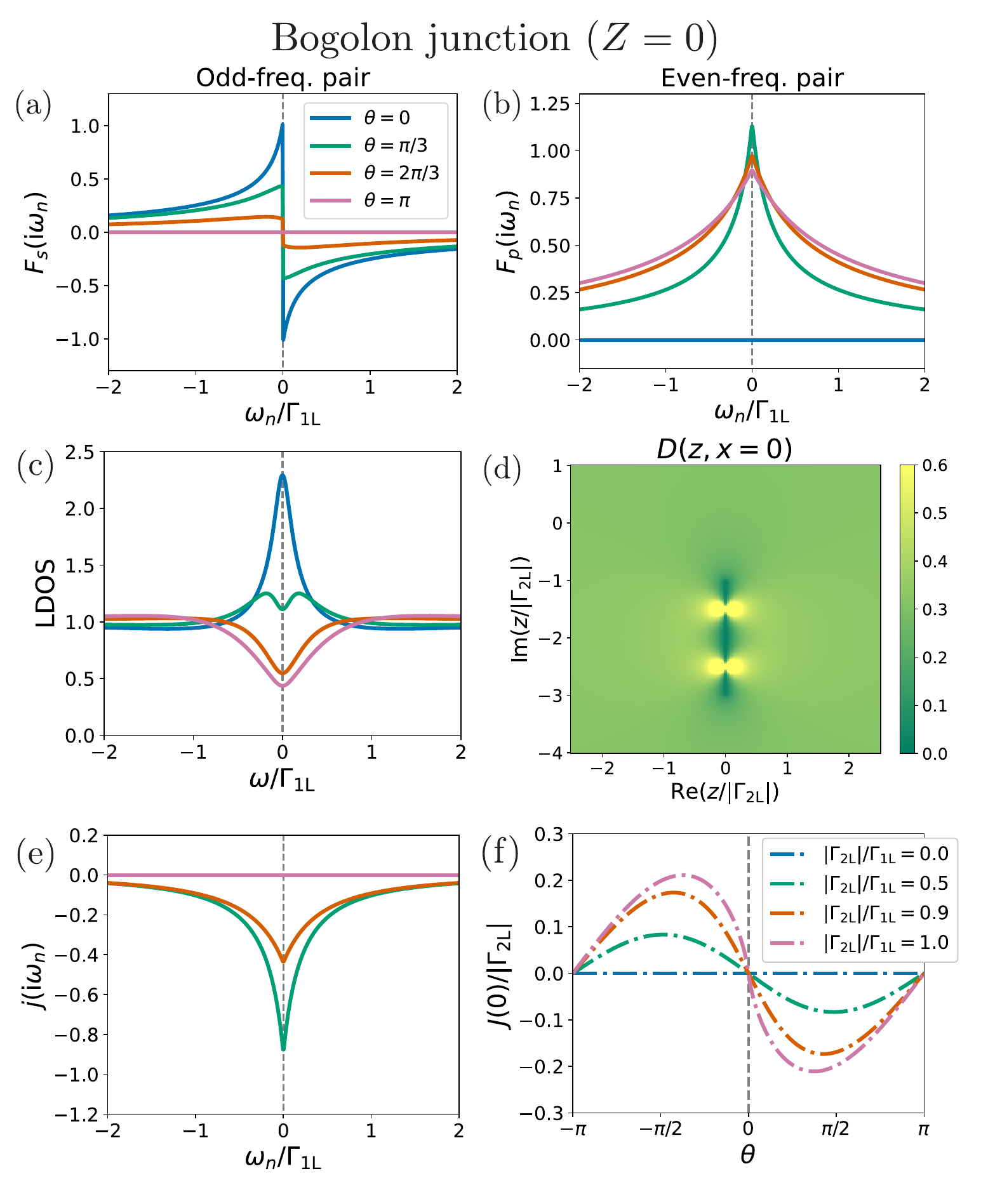}
    \caption{
    Physical quantities at $x = 0$ for bogolon junction without barrier potential.
    (a) The $s$-wave component of the pair amplitude, 
    (b) the $p$-wave component of the pair amplitude,
    (c) the LDOS of bogolons normalized by its value of clean limit ($\Gamma_{1\mrm L} = \Gamma_{2\mrm L} = 0$), 
    (d) the LDOS of bogolons in the complex energy plane for $\theta = 2\pi/3$, which is defined by the extension of $\om$ to the complex plane $z$, i.e., $D(\om, x=0) \to D(z, x=0)$, 
    (e) $j(\imu\omega_{n}) = j(\imu\omega_{n}, x = 0)$,
    and (f) $J(x=0) = (1/\be)\sum_n j(\imu\om_n)$.
    Figure legends of (b)-(d) are the same as that of (a).
    We set $|\Gamma_{2\mrm L}|/\Gamma_{1\mrm L}=0.9$ in (a)--(e).
    }
    \label{fig:mcmillan_bog1}
\end{figure}

In contrast to the semi-infinite case discussed in Sec.~\ref{sec:semi_infinite}, the pair amplitude is the mixed function of even- and odd-frequencies.
The $s$-wave component of the pair amplitude defined by Eq.~\eqref{eq:fs_def} is given by
\begin{align}
    &F_s(\imu\om_n, x) 
    = \frac{-\imu\Gamma_{2\mrm L} \sgn\om_n m}{\imu k_{\mrm F} \hbar^2 \Omega_{\mrm L}(\imu\om_n)}
    \Bigg[ 1
    - \epn^{-\imu (k_{\mrm L}^+ - k_{\mrm L}^-) x} \nt
    &\hspace{-20pt}
    \times\frac{(\imu\om_n + \imu\Gamma_{1\mrm L}\sgn\om_n)^2\sin^2(\theta/2) + \imu\Omega_{\mrm L}(\imu\om_n)^2 \sin(\theta/2) \cos(\theta/2)}{(\imu\om_n + \imu\Gamma_{1\mrm L})^2 + |\Gamma_{2\mrm L}|^2 \cos^2(\theta/2)} 
    \Bigg]. \label{eq:fs_bog}
\end{align}
The first term is the contribution from bulk and the second one is from the Andreev reflection.
The numerical result of $F_s(\imu\om_n, x = 0)$ is shown in Fig.~\ref{fig:mcmillan_bog1} (a).
The phase of $F_s(\imu\om_n, x = 0)$ is independent of $\om_n$, which is the same feature as the $s$-wave SC junction. 
We choose the phase such that $F_s(\imu\om_n, x = 0)$ becomes real for the plot [this choice also applies to Figs.~\ref{fig:mcmillan_bog1} (b), \ref{fig:mcmillan_bog2} (a), and \ref{fig:mcmillan_bog2} (b)].
The odd-frequency dependence of $F_s(\imu\om_n, x)$ can be directly verified from Eq.~\eqref{eq:fs_bog} by noting that $\Omega_{\mathrm{L}}(\imu\omega_n)$ is an even function of $\om_n$.
The height of the pair amplitude gradually decreases as the relative phase $\theta$ increases and reaches zero at $\theta = \pi$.

In the present bogolon junction, the even-frequency pairing is induced near the interface due to the translational symmetry breaking.
To see this, we evaluate the $p$-wave component of the pair amplitude defined by Eq.~\eqref{eq:fp_def} as
\begin{align}
    &F_p(\imu\om_n, x)
    = \frac{2\Gamma_{2\mrm L}\sgn\om_n m}{\hbar^2} \nt
    &\times\frac{ (\imu\om_n + \imu\Gamma_{1\mrm L}\sgn\om_n) \sin(\theta/2) \epn^{-\imu\theta/2}}{(\imu\om_n + \imu\Gamma_{1\mrm L} \sgn\om_n)^2 + |\Gamma_{2\mrm L}|^2 \cos^2(\theta/2)} 
    \epn^{-\imu (k_{\mrm L}^+ - k_{\mrm L}^-) x}.
    \label{eq:fp_bog}
\end{align}
In Eq.~\eqref{eq:fp_bog}, the bulk terms vanish, and then only the term from Andreev reflection contributes.
The results at $x = 0$ is shown in Fig.~\ref{fig:mcmillan_bog1} (b).
In contrast to the $s$-wave SC, the pair amplitude of bogolons does not diverge near $\theta = \pi$ due to the presence of finite $\Gamma_{1\mrm L}$ in the denominator in the second line of Eq.~\eqref{eq:fp_bog}.

\subsubsection{Diagonal quantities}

Now we turn to the diagonal quantities such as the LDOS and $J(x)$.
The specific form of the LDOS of bogolons is given by
\begin{align}
    &D(\om, x)
    = D_{\mrm{bulk}}(\om) \nt
    &+ \frac{2m}{k_{\mrm F} \hbar^2 \pi}\real\bigg[ \frac{\om + \imu\Gamma_{1\mrm L}}{\Omega_{\mrm{ret}}(\omega)}
    \frac{|\Gamma_{2\mrm L}|^2 \sin^2(\theta/2) \epn^{-\imu(k_{\mrm L}^+ - k_{\mrm L}^-)x}}{(\om + \imu\Gamma_{1\mrm L})^2 + |\Gamma_{2\mrm L}|^2 \cos^2(\theta/2)} \bigg], \label{eq:ldos_bog}
\end{align}
where $D_{\mrm{bulk}}(\om)$ is a bulk DOS of bogolons defined by
\begin{align}
    D_{\mrm{bulk}}(\om) = \frac{2m}{k_{\mrm F}\hbar^2 \pi}\real \frac{\om + \imu\Gamma_{1\mrm L}}{\Omega_{\mrm{ret}}(\omega)} \label{eq:bulkdos}
\end{align}
with $\Omega_{\mrm{ret}}(\omega) = \Omega(\om + \imu 0^+) \sgn\om$.
The second line of Eq.~\eqref{eq:ldos_bog} is the Andreev reflection part, which takes a finite value in the non-uniform case ($\theta \neq 0$).
We list in Table~\ref{tab:comparison} the expressions of the LDOS of bogolons in the low-$\omega$ limit for $\theta = 0$ [row (c)] and $\theta = \pi$ [row (d)].

Figure~\ref{fig:mcmillan_bog1} (c) shows the LDOS of bogolons at $x=0$.
In the uniform case ($\theta = 0$), the LDOS has the zero-energy peak, which is consistent with the previous calculations in bulk \cite{Miki21, Miki24}.
As $\theta$ increases, the pseudo-gap appears at zero energy.
The gap formation in the LDOS is incomplete for any $\theta$, i.e., the value of the LDOS at $\omega=0$ is always finite.

As discussed above, the LDOS behaviors of $s$-wave SC and bogolon junctions are quite different.
Nevertheless, the correspondence between the two junctions can be visualized by considering the LDOS in a complex frequency space.
Let us extend $\om$ in the LDOS of bogolons onto the complex plane as $D(\om, x) \to D(z, x)$ with $z\in\mathbb{C}$.
Figure~\ref{fig:mcmillan_bog1} (d) shows the LDOS in the complex plane at $x = 0$ for the bogolon junction.
The positions of the peaks are given by $\om_{\mrm{ABS}}(\theta) = -\imu\Gamma_{1\mrm L} \pm \imu |\Gamma_{2\mrm L}|\sqrt{1 - \sin^2(\theta/2)}$ located on the imaginary axis.
On the other hand, for the $s$-wave SC case, these peak positions appear on the real axis as discussed in Sec.~\ref{sec:swave}.
The LDOS of bogolons [Fig.~\ref{fig:mcmillan_bog1} (c)] corresponds to the $s$-wave SC junction with $90^\circ$-rotation in the complex plane [Fig.~\ref{fig:mcmillan_s1} (d)].

Next, we show the results for quasiparticle current $J(x)$.
We first consider the contribution at $\omega_n$ defined by Eq.~\eqref{eq:jom_def}:
\begin{align}
    &j(\imu\om_n, x) 
    = \frac{2m|\Gamma_{2 \mrm L}|^2}{\hbar^2} \nt
    &\times\frac{\sin(\theta/2) \cos(\theta/2) \epn^{-\imu (k_{\mrm L}^+ - k_{\mrm L}^-) x}}{(\imu\om_n + \imu\Gamma_{1\mrm L}\sgn\om_n)^2 + |\Gamma_{2\mrm L}|^2 \cos^2(\theta/2)}. \label{eq:jom_bog}
\end{align}
Here, Andreev reflection only contributes to $j(\imu\om_n,x)$ similar to Eq.~\eqref{eq:fp_bog}.
The frequency dependence of Eq.~\eqref{eq:jom_bog} at $x = 0$ is shown in Fig.~\ref{fig:mcmillan_bog1} (e).
$j(\imu\om_n,x = 0)$ becomes zero at $\theta = 0$ or $\pi$.
The value at $\om = 0$ for $\theta = 2\pi/3$ case is larger than the value at $\om = 0$ for $\theta = \pi/3$ case, which is in contrast to the $s$-wave SC junction. 
The absolute value of $j(\imu\om_n, x)$ for $\om_n \to 0$ takes the maximum value when $\theta = \cos^{-1}(\Gamma_{1\mrm L}^2/(2\Gamma_{1\mrm L}^2 - |\Gamma_{2\mrm L}|^2))$.

For zero-temperature limit $\be \to \infty$, we can take the summation of $\om_n$:
\begin{align}
    J(x = 0) = -\frac{m|\Gamma_{2\mrm L}|}{\pi\hbar^2} \sin\frac{\theta}{2} \ln\bigg(\frac{\Gamma_{1\mrm L} + |\Gamma_{2\mrm L}|\cos(\theta/2)}{\Gamma_{1\mrm L} - |\Gamma_{2\mrm L}|\cos(\theta/2)}\bigg). \label{eq:j_ln}
\end{align}
Compared with the $s$-wave SC junction case, the $\theta$-dependence of Eq.~\eqref{eq:j_ln} has the additional factor $\displaystyle \ln\bigg(\frac{\Gamma_{1\mrm L} + |\Gamma_{2\mrm L}|\cos(\theta/2)}{\Gamma_{1\mrm L} - |\Gamma_{2\mrm L}|\cos(\theta/2)}\bigg)$ (see row (e) in Table~\ref{tab:comparison}).
Due to this logarithmic factor, $J(x=0)$ is a continuous function even at $\theta=\pi$, which is a clear difference from the $s$-wave SC junction.
Additionally, in comparison to the $s$-wave SC junction case in Eq.~\eqref{eq:j_s_t0}, the different sign for the bogolon junction is a consequence of odd-frequency pair potential, which is referred to as the $\pi$-junction. 
Namely, the relation $[\hat \Sigma(-\imu\om_n, x)]_{12} = -[\hat \Sigma(\imu\om_n, x)]_{12}$, is the origin of the minus sign.
Figure~\ref{fig:mcmillan_bog1} (f) shows $J(x=0)$ for several values of $|\Gamma_{2\mrm L}|/\Gamma_{1\mrm L}$.
The maximum value of $|J(x=0)|$ shifts towards $\theta = 0$ as $|\Gamma_{2\mrm L}|/\Gamma_{1\mrm L}$ increases because the contribution of the factor $\cos(\theta/2)$ in Eq.~\eqref{eq:j_ln} becomes larger.

\subsubsection{Correlation between LDOS and pair amplitudes}

\begin{figure}[tbp]
    \centering
    \includegraphics[width=8.6cm]{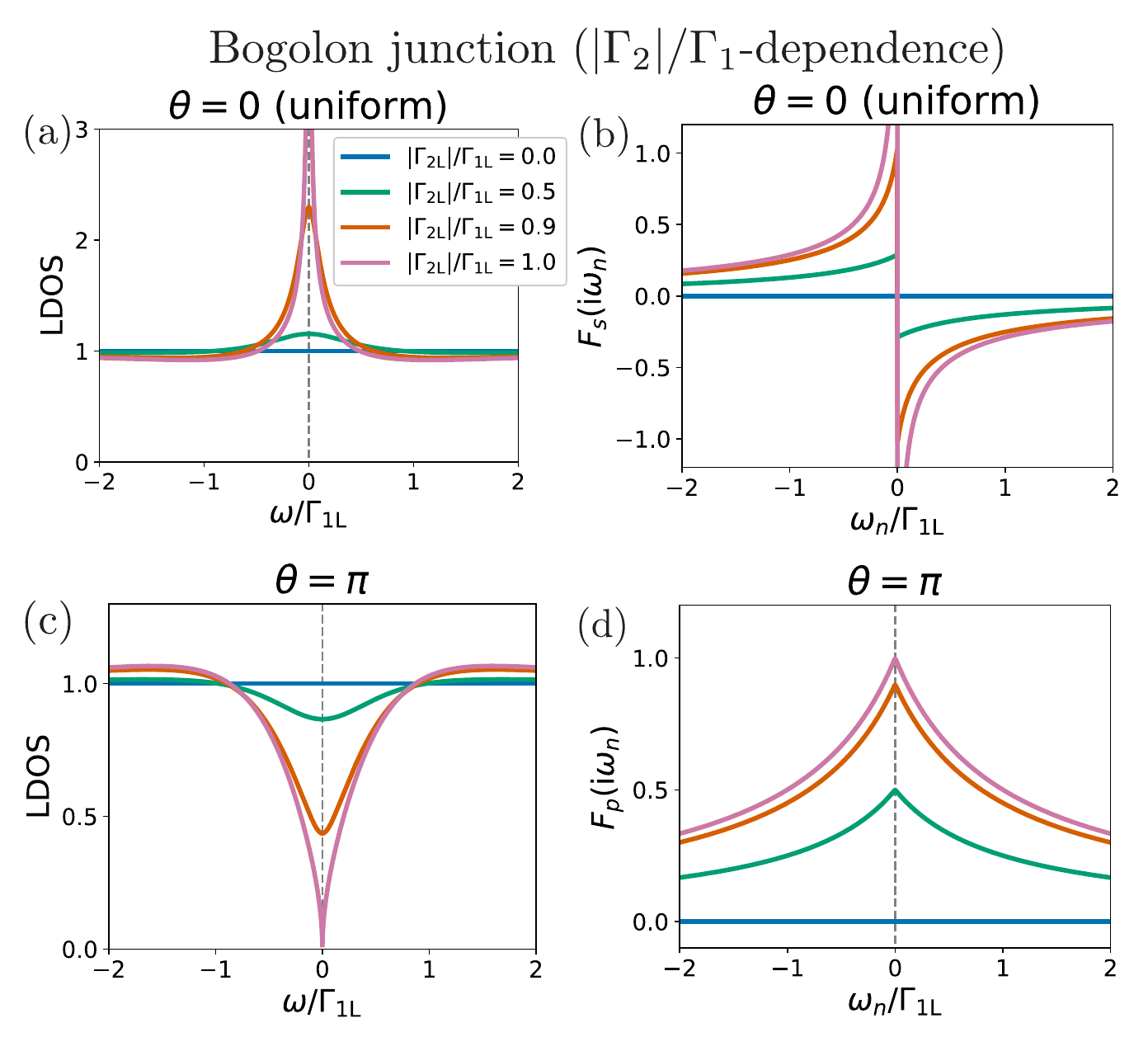}
    \caption{
    Frequency-dependence of LDOS of bogolons normalized by its value of clean limit ($\Gamma_{1\mrm L} = \Gamma_{2\mrm L} = 0$) and pair amplitudes at $Z = 0$.
    (a) LDOS at $\theta = 0$ and 
    (b) $s$-wave component of pair amplitude at $\theta = 0$.
    (c) LDOS at $\theta = \pi$ and
    (d) $p$-wave component of pair amplitude at $\theta = \pi$.
    The lines in (b)-(d) are shared with those in (a).
    }
    \label{fig:mcmillan_bog2}
\end{figure}

Let us discuss the correlation between the LDOS of bogolon and the pair amplitudes $F_s(\imu\om_n, x)$ and $F_p(\imu\om_n, x)$.
We here focus on the $|\Gamma_{2\mrm L}|/\Gamma_{1\mrm L}$-dependence for the uniform case ($\theta = 0$) and the non-uniform case ($\theta = \pi$).
Since the relation $\Gamma_{1\mrm L} > |\Gamma_{2\mrm L}|$ must be satisfied to guarantee the positive DOS, we consider the case for $0 < |\Gamma_{2\mrm L}|/\Gamma_{1\mrm L} < 1$.

First, in the uniform case ($\theta = 0$), the LDOS and $s$-wave pair amplitude are shown in Fig.~\ref{fig:mcmillan_bog2} (a) and (b), respectively.
Note that the $p$-wave component of pair amplitude is zero in this uniform case. 
The LDOS at $\omega = 0$ in (a) increases as $|F_s(\imu\omega_n \to 0, x=0)|$ in (b) increases. 
Especially in the limit of $|\Gamma_{2\mrm L}|/\Gamma_{1\mrm L} \to 1$, both LDOS at $\om = 0$ and $F_s(\imu\omega_n \to 0, x=0)$ diverge.
We will revisit the relation between the LDOS and $F_s$ in Sec.~\ref{sec:quasiclassical}.

Second, in the $\theta = \pi$ case, the LDOS and $p$-wave pair amplitude are shown in Fig.~\ref{fig:mcmillan_bog2} (c) and (d), respectively.
The $s$-wave component of pair amplitude is zero in this case. 
The depth of the pseudogap of the LDOS in (c) becomes larger as $F_p(\imu\omega_n, x=0)$ in (d) increases.
Hence, the depth of pseudogap in the LDOS is correlated with the magnitude of the $p$-wave pair amplitude.

Therefore, the zero-energy LDOS is correlated with pair amplitudes.

\subsection{Effect of barrier potential \label{sec:barrier}}

\begin{figure}[tbp]
    \centering
    \includegraphics[width=8.6cm]{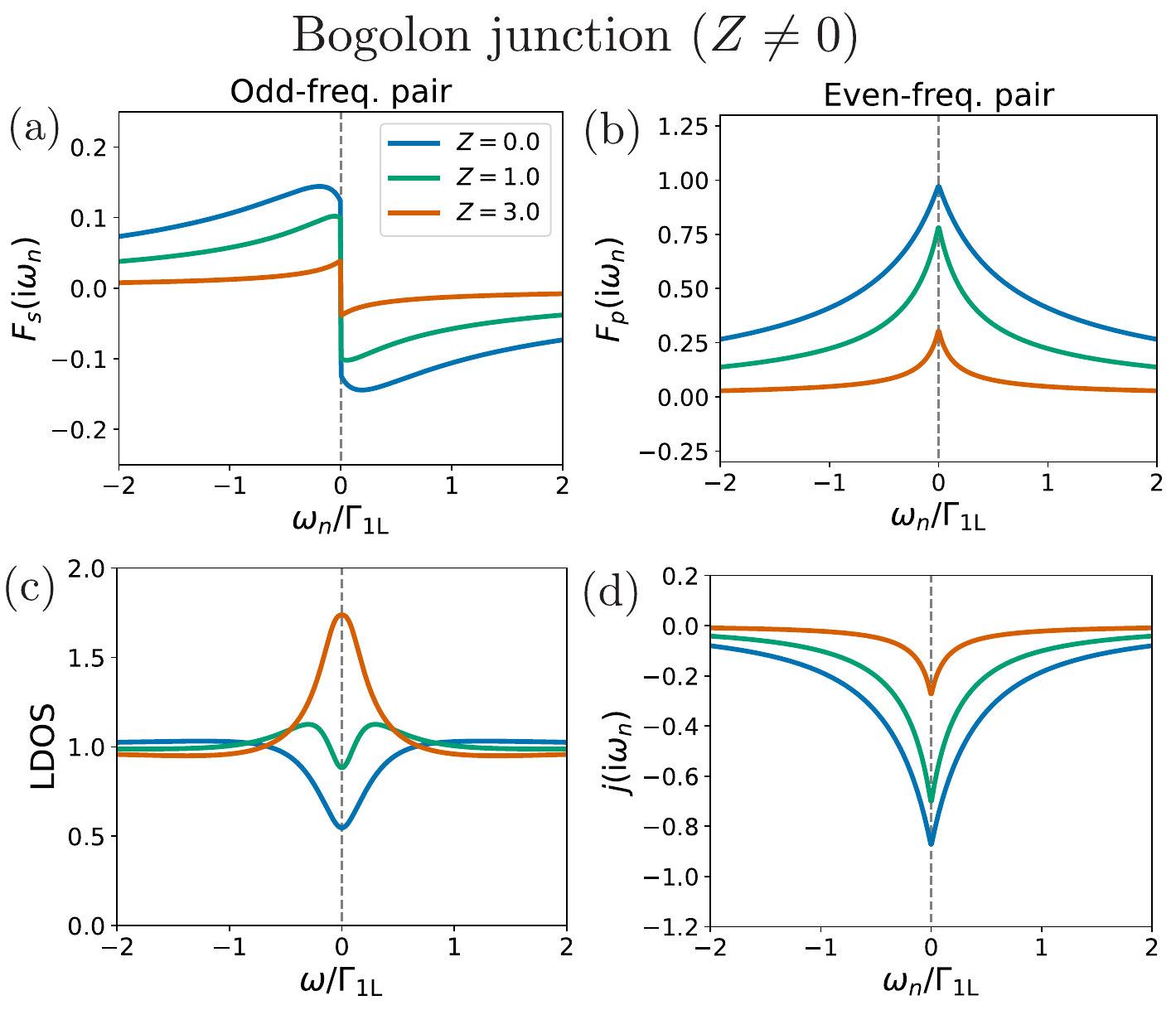}
    \caption{
    Frequency-dependence of physical quantities of bogolon junction at $x = 0$ for several values of $Z$ with $|\Gamma_{2\mrm L}|/\Gamma_{1\mrm L}=0.9$.
    (a) The $s$-wave component pair amplitude, 
    (b) the $p$-wave component pair amplitude, 
    (c) the LDOS of bogolons normalized by its value of clean limit ($\Gamma_{1\mrm L} = \Gamma_{2\mrm L} = 0$), 
    and (d) $j(\imu\om_n)$.
    The lines in (b)-(d) are shared with those in (a).
    The relative phase is chosen as $\theta = 2\pi/3$.
    }
    \label{fig:mcmillan_bog3}
\end{figure}

In this subsection, we study the effect of the barrier potential.
The barrier-potential dependence of the Green's function, as expressed in Eqs.~\eqref{eq:mcmillan_bog}--\eqref{eq:b} [see also Eqs.~\eqref{eq:mcmillan_s}--\eqref{eq:b_s}], is controlled by the parameter $Z$ defined in Eq.~\eqref{eq:z}.
In the limit of $Z \to \infty$, $Z$-dependence of $\tilde a_{\mrm{out}}^{(\pm)}$ given by Eq.~\eqref{eq:a} is roughly expressed as $\tilde a_{\mrm{out}}^{(\pm)} \sim 1/Z^2$, while $\tilde b_{\mrm{out}}^{(\pm)}$ is expressed as $\tilde b_{\mrm{out}}^{(\pm)} \sim 1$.
Thus, the contribution of Andreev reflection becomes smaller than the normal reflection.

Figure~\ref{fig:mcmillan_bog3} shows the physical quantities of bogolons at $x = 0$.
The $s$-wave pair amplitude, $p$-wave pair amplitude, the LDOS of bogolons, and $j(i\omega_{n})$ are shown in  
(a), (b), (c), and (d), respectively. 
The relative phase is chosen as $\theta = 2\pi/3$.
The absolute values of the pair amplitudes at $x = 0$ in (a) and (b) become smaller as $Z$ increases due to high barrier potential.

\begin{table*}[tb]
    \centering
    \begin{tabular}{c|c||c|c}
        \hline
        &
        & Bogolon junction & $s$-wave SC junction \\ \hline\hline
        (a)&
        Bulk pair ($s$-wave) & Odd-freq. pair & Even-freq. pair \\ \hline
        (b)&
        Induced pair at the interface ($p$-wave) & Even-freq. pair & Odd-freq. pair \\ 
        \hline
        (c)&
        \begin{tabular}{l}
            Zero-energy LDOS \\
            at $(\theta,Z)=(0,0)$ (bulk) [$D_{\mrm{bulk}}(\om = 0)$]
        \end{tabular}
        & 
        $\displaystyle \frac{2m}{k_{\mrm F}\hbar^2 \pi} 
        \frac{1}{\sqrt{1 - (|\Gamma_{2\mrm L}|/\Gamma_{1\mrm L})^2}}
        $
        & $0$ (gapped) \\ \hline
        (d)&
        \begin{tabular}{l}
            Zero-energy LDOS \\
            at $(\theta,Z)=(\pi,0)$  [$D(\om = 0, x = 0)$]
        \end{tabular}
        & $\displaystyle \frac{2m}{k_{\mrm F}\hbar^2 \pi} \sqrt{1 - (|\Gamma_{2\mrm L}|/\Gamma_{1\mrm L})^2}$ & $\displaystyle \frac{2m |\Delta_{\mrm L}|}{k_{\mrm F}\hbar^2 \pi} \delta(\om)$
        \\[2mm] \hline
        (e)&
        \begin{tabular}{l}
            Current phase relation for $-\pi < \theta < \pi$ \\
            at $T=0$ ($Z=0$) [$J(x=0)$]
        \end{tabular}
        & $\displaystyle -\frac{m|\Gamma_{2\mrm L}|}{\pi\hbar^2} \sin\frac{\theta}{2} \ln\bigg(\frac{\Gamma_{1\mrm L} + |\Gamma_{2\mrm L}|\cos(\theta/2)}{\Gamma_{1\mrm L} - |\Gamma_{2\mrm L}|\cos(\theta/2)}\bigg)$ & $\displaystyle \frac{2m|\Delta_{\mrm L}|}{\hbar^2} \sin\frac{\theta}{2}$
        \\
        \hline
        (f)&
        \begin{tabular}{l}
            Current phase relation for $-\pi < \theta < \pi$ \\
            at $T=0$ ($Z\to \infty$) [$J(x=0)$]
        \end{tabular}
        & $\displaystyle -\frac{m|\Gamma_{2\mrm L}|}{2\pi\hbar^2 Z^2}
        \ln\bigg(\frac{\Gamma_{1\mrm L} + |\Gamma_{2\mrm L}|}{\Gamma_{1\mrm L} - |\Gamma_{2\mrm L}|}\bigg) \sin\theta$
        & $\displaystyle \frac{m|\Delta_{\mrm L}|}{\hbar^2 Z^2} \sin\theta$
        \\
        \hline
    \end{tabular}
    \caption{
    Comparison of functional forms between bogolon junction (the third column) and $s$-wave SC junction (the fourth column).
    We show the frequency dependence of pair amplitudes in (a) bulk and (b) interface.
    The zero-energy LDOS is shown for (c) $\theta = 0$ and (d) $\theta=\pi$ at $x = 0$, and the quasiparticle current $J(x=0)$ for (e) $Z = 0$ and (f) $Z\to\infty$ at zero temperature.
    }
    \label{tab:comparison}
\end{table*}

Let us take a closer look at the LDOS of bogolons, which is given by
\begin{align}
    &D(\om, x = 0) \nt
    &= \frac{2m}{\pi k_{\mrm F}\hbar^2} \real \frac{(\om + \imu\Gamma_{1\mrm L}) \Omega_{\mrm{ret}}(\omega)}{Z^2 \Omega_{\mrm{ret}}(\omega)^2 + (\om + \imu\Gamma_{1\mrm L})^2 + |\Gamma_{2\mrm L}|^2 \cos^2(\theta/2)}. \label{eq:ldos_bog_z}
\end{align}
Figure~\ref{fig:mcmillan_bog3} (c) shows the LDOS at $x = 0$.
The pseudo-gap for a small value of $Z$ [$Z=0$ and $1$ in Fig.~\ref{fig:mcmillan_bog3} (c)] changes into the zero energy peak for a larger value of $Z$ [$Z=3$ in Fig.~\ref{fig:mcmillan_bog3} (c)], which is proportional to the bulk value as shown in Eq.~\eqref{eq:ldos_bog_zinf}.
Indeed, in the limit of $Z \to \infty$, we obtain
\begin{align}
    D(\om, x = 0) 
    &= \frac{2m}{\pi Z^2 k_{\mrm F}\hbar^2} \real \frac{\om + \imu\Gamma_{1\mrm L}}{\Omega_{\mrm{ret}}(\omega)} \nt
    &= \frac{1}{Z^2} D_{\mrm{bulk}}(\om, x = 0), \label{eq:ldos_bog_zinf}
\end{align}
where $D_{\mrm{bulk}}(\om)$ is the bulk DOS defined in Eq.~\eqref{eq:bulkdos}.
In this limit, Eq.~\eqref{eq:ldos_bog_zinf} is independent of $\theta$ and proportional to the bulk DOS and decays with the factor $1/Z^2$.
The appearance of the bulk DOS in $Z \to \infty$ limit is a common feature with the $s$-wave SC junction [see Eq.~\eqref{eq:ldos_s_zinf}].

Let us proceed to consider $J(x = 0)$, which is given by the frequency summation of Fig.~\ref{fig:mcmillan_bog3} (d).
To see the $Z$-dependence, we evaluate $J(x = 0)$ for zero-temperature limit in the analytical expression:
\begin{align}
    &J(x = 0) 
    = -\frac{m|\Gamma_{2\mrm L}|}{2\pi\hbar^2} \frac{1}{Z^2 + 1}\sqrt{\frac{Z^2 + 1}{Z^2 + \cos^2(\theta/2)}} \nt 
    &\times \sin\theta \ln\Bigg(\frac{\Gamma_{1\mrm L}\sqrt{Z^2 + 1} + |\Gamma_{2\mrm L}|\sqrt{Z^2 + \cos^2(\theta/2)}}{\Gamma_{1\mrm L}\sqrt{Z^2 + 1} - |\Gamma_{2\mrm L}|\sqrt{Z^2 + \cos^2(\theta/2)}}\Bigg). \label{eq:j_bog_z}
\end{align}
Specifically in the limit of $Z \to \infty$, we obtain
\begin{align}
    J(x = 0) 
    &= -\frac{m|\Gamma_{2\mrm L}|}{2\pi\hbar^2 Z^2}  
    \sin\theta \ln\bigg(\frac{\Gamma_{1\mrm L} + |\Gamma_{2\mrm L}|}{\Gamma_{1\mrm L} - |\Gamma_{2\mrm L}|}\bigg).
\end{align}
The $\theta$-dependence is determined by the factor $\sin\theta$, and $J(x = 0)$ is proportional to $1/Z^2$ in the limit of $\be \to \infty$.

\subsection{Comparison between bogolon junction and $s$-wave SC junction}

We have studied the bogolon junction with the bulk odd-frequency pair and made a contrast against the conventional $s$-wave SC with bulk even-frequency pair.
We summarize the main results in Table~\ref{tab:comparison}.
The third column shows the properties of the bogolon junction, and the fourth column shows those of the $s$-wave SC junction.

At the interface, the even-frequency $p$-wave pair is induced by the translational symmetry breaking for the bogolon junction, and the odd-frequency $p$-wave pair is induced for the $s$-wave SC junction [rows (a) and (b)].

The bogolon junction has a zero-energy peak in the LDOS in bulk ($\theta = 0$), whose height is given by $\sim [1 - (|\Gamma_{2\mrm L}|/\Gamma_{1\mrm L})^2]^{-1/2}$, while the $s$-wave SC has a superconducting gap [row (c)].
For $\theta = \pi$, the LDOS of the bogolon junction at the interface has a pseudo-gap with the depth of $\sim [1 - (|\Gamma_{2\mrm L}|/\Gamma_{1\mrm L})^2]^{1/2}$, which is the inverse of the bulk DOS, while the $s$-wave SC has a delta-function behavior due to the Andreev bound state [row (d)].

The rows (e) and (f) show the current phase relations of the two cases, which have opposite signs due to the different frequency dependence of the pair potential.
For $Z = 0$ shown in row (e), the bogolon junction includes an additional logarithmic factor compared to the $s$-wave SC junction.
The $\theta$-dependence becomes the same for both the junctions in $Z \to \infty$ limit as shown in row (f), except for the presence of a minus sign in the bogolon junction.

In this way, the bogolon junction has the characteristic features compared with the conventional $s$-wave SC junction.

\section{Quasiclassical Green's function \label{sec:quasiclassical}}

In the above, we have mainly focused on the Green's function at $x = 0$.
On the other hand, since we consider the non-uniform system, the unique properties of bogolons can also be captured in the $x$ dependence of the Green's function.
While the McMillan Green's function includes the rapidly oscillating components with a characteristic scale of $k_{\mrm F}^{-1}$, we focus on the quasiclassical Green's function with slowly-varying spatial components.

We construct the quasiclassical Green's function by combining the method in Refs.~\cite{Ashida89, Nagato93} and McMillan Green's function derived in Sec.~\ref{sec:mcmillan}.
We start by decomposing the Green's function as
\begin{align}
    \hat G(\imu\om_n, x, x') 
    = \sum_{\al \al' = \pm} \hat G^{\al \al'}(\imu\om_n, x, x') \epn^{\imu k_{\mrm F}(\al x - \al' x')}. \label{eq:g_decomp}
\end{align}
The $x$ dependence of each component $\hat G^{\al \al'}(\imu\om_n, x, x')$ does not include rapidly oscillating contribution with a characteristic scale of $k_{\mrm F}^{-1}$, which is consistent with the picture of the quasiclassical Green's function.
However, $\hat{G}^{\alpha\alpha'}(\imu\om_n,x,x')$ is discontinuous at $x = x'$. 
Therefore, to satisfy the Eilenberger equation \cite{Kopnin_book}, we define the continuous semiclassical Green's function as
\begin{align}
    \hat g^{\pm\pm}(\imu\om_n, x)
    &= \pm \imu \hat 1 - 2v_{\mrm F} \hat \tau^z \hat G^{\pm\pm}(\imu\om_n, x - 0^+, x), \label{eq:quasiclassical}
\end{align}
where $v_{\mrm F} = \hbar k_{\mrm F}/m$ is a Fermi velocity \cite{Ashida89, Nagato93}.

Since the overall phase does not change the physical properties, we fix the phase on the left side as $\theta_{\mrm L} = 0$.
Then, from Eq~\eqref{eq:quasiclassical}, we obtain the quasiclassical Green's function for $x < 0$ as
\begin{widetext}
    \begin{align}
    \hat g^{\pm\pm}(\imu\om_n, x)
    &= \frac{\imu}{\Omega_{\mrm L}(\imu\om_n)} \Big[ 
    \Gamma_{2\mrm L}\sgn\om_n \hat \tau^y 
    + (\imu\om_n + \imu\Gamma_{1\mrm L}\sgn\om_n) \hat \tau^z \nt
    &+ \tilde a_{\mrm{out}}^{(\mp)} \epn^{-\imu k_{\mrm F}(\Omega_{\mrm L}(\imu\om_n)/\mu) x} \Big(
    \pm\Omega_{\mrm L}(\imu\om_n) \hat \tau^x 
    - \imu(\imu\om_n + \imu\Gamma_{1\mrm L}\sgn\om_n) \hat \tau^y 
    + \imu|\Gamma_{2\mrm L}|\sgn\om_n \hat \tau^z \Big) \Big]. \label{eq:quasi1}
\end{align}
\end{widetext}
The first line is a bulk part, and the second line, which is proportional to $\tilde a_{\mrm{out}}^{(\mp)}$, describes the contributions from the Andreev reflection.
Using Eq.~\eqref{eq:quasi1}, we can evaluate the physical quantities defined in Sec.~\ref{sec:quantities} in the quasiclassical representations, whose specific forms are listed in Appendix~\ref{sec:app_quasiclassical}.
One can confirm that Eq.~\eqref{eq:quasi1} 
satisfies the following Eilenberger equation \cite{Ashida89, Nagato93, TextTanaka2021}:
\begin{align}
    -\imu v_{\mrm F} \frac{\del}{\del x} \hat g^{\pm\pm}(\imu\om_n, x) 
    &= \pm(\imu\om_n\hat 1 - \hat \Sigma)\hat \tau^z \hat g^{\pm\pm}(\imu\om_n, x) \nt 
    &\mp \hat g^{\pm\pm}(\imu\om_n, x) (\imu\om_n\hat 1 - \hat \Sigma)\hat \tau^z.
\end{align}
We can also check that $\hat g^{\pm\pm}(\imu\om_n, x)$ satisfies the normalization condition
\begin{align}
    \hat g^{\pm\pm}(\imu\om_n, x)^2 = -\hat 1.
\end{align}

One can extract the factor of spatial dependence from Eq.~\eqref{eq:quasi1} as
\begin{align}
    &\epn^{-\imu k_{\mrm F}(\Omega_{\mrm L}(\imu\om_n)/\mu) x} \nt
    &= \exp(2\sqrt{(\om_n + \Gamma_{1\mrm L}\sgn\om_n)^2 - |\Gamma_{2\mrm L}|^2} x /\hbar v_{\mrm F}). \label{eq:exp}
\end{align}
Since $\sqrt{(\om_n + \Gamma_{1\mrm L}\sgn\om_n)^2 - |\Gamma_{2\mrm L}|^2}$ is real, Eq.~\eqref{eq:exp} does not exhibit an oscillating part and is suppressed with increasing distance from the interface (The $s$-wave SC junction has similar feature because $\sqrt{\om_n^2 + |\Delta_{\mrm L}|^2}$ in Eqs.~\eqref{eq:exp_s} is also real.).
Then, the Green's function converges to the bulk value for the limit of $x \to -\infty$.

On the other hand, the spatial dependence of the retarded Green's function is extracted as
\begin{align}
    \exp(-2\imu\sqrt{(\om + \imu\Gamma_{1\mrm L})^2 + |\Gamma_{2\mrm L}|^2} \ \sgn\om \  x / \hbar v_{\mrm F}). \label{eq:exp_ret}
\end{align}
In contrast to Matsubara form, Eq.~\eqref{eq:exp}, the above quantity includes both damping and oscillating parts, because $\sqrt{(\om + \imu\Gamma_{1\mrm L})^2 + |\Gamma_{2\mrm L}|^2} \sgn\om$ with $\Gamma_{1\mrm L} > 0$ exhibits both real and imaginary part.
Especially for small energy $[\om \ll(\Gamma_{1\mrm L}^2-|\Gamma_{2\mrm L}|^2)/\Gamma_{1\mrm L}]$, the oscillating part of Eq.~\eqref{eq:exp_ret} reduces to $\exp(-2\imu\Gamma_{1\mrm L} |\om| x / \hbar v_{\mrm F} \sqrt{\Gamma_{1\mrm L}^2 - |\Gamma_{2\mrm L}|^2})$.
Namely, the quasiclassical Green's function has oscillating components with the frequency-dependent length scale $k_{\mrm F}^{-1} \sqrt{\Gamma_1^2 - |\Gamma_{2\mrm L}|^2} \mu /\Gamma_{1\mrm L}|\om|$, even though we have excluded rapidly oscillating components with a period of $k_{\mrm F}^{-1}$.

In the $s$-wave SC case [Eq.~\eqref{eq:exp_s_ret}], $\sqrt{\om^2 - |\Delta_{\mrm L}|^2}\sgn\om$ is real for $|\om| > |\Delta_{\mrm L}|$ and purely imaginary for $|\om| < |\Delta_{\mrm L}|$.
This indicates that the LDOS has oscillation without dumping in $|\om| > |\Delta_{\mrm L}|$ while has dumping without oscillation in $|\om| < |\Delta_{\mrm L}|$.

We define the characteristic length by taking the limit of $\om \to 0$ in Eq.~\eqref{eq:exp_ret}, which is expressed as $\xi = \hbar v_{\mrm F}/\sqrt{\Gamma_{1\mrm L}^2 - |\Gamma_{2\mrm L}|^2}$.
We have evaluated a similar quantity for the bulk in Ref.~\cite{Miki21}, which is defined by the relative position of two bogolons and hence corresponds to the pair radius.
In contrast, the length $\xi$ in this paper is defined by the center of mass coordinate of Green's function, which corresponds to the coherent length.
Nevertheless, we obtain the same form for the two length scales.

Finally, we comment on the correlation between the LDOS and the pair amplitude of bogolon in bulk limit.
The second line of Eq.~\eqref{eq:quasi1} for $\theta = 0$ is zero.
Then, Eq.~\eqref{eq:quasi1} reduces to
\begin{align}
    &\hat g^{\pm\pm}(\imu\om_n, x) = f(\imu\om_n, x) \hat \tau^y + g(\imu\om_n, x) \hat \tau^z
\end{align}
with
\begin{align}
    &f(\imu\om_n, x) = \frac{\Gamma_{2\mrm L}\sgn\om_n}{\sqrt{(\om_n + \Gamma_{1\mrm L})^2 - |\Gamma_{2\mrm L}|^2}}, \\
    &g(\imu\om_n, x) = \frac{\imu\om_n + \imu\Gamma_{1\mrm L}\sgn\om_n}{\sqrt{(\om_n + \Gamma_{1\mrm L})^2 - |\Gamma_{2\mrm L}|^2}}.
\end{align}
Note that $g(\imu\om_n, x)$ is purely imaginary, and $f(\imu\om_n, x)$ is real, which is a consequence of odd-frequency pairing.
In this case, the normalization condition is given by
\begin{align}
    f(\imu\om_n, x)^2 + g(\imu\om_n, x)^2 = -1. \label{eq:quasi_norm}
\end{align}
Since $f(\imu\om_n, x)$ is real and $g(\imu\om_n, x)$ is purely imaginary, if $|f(\imu\om_n, x)|$ increases, $|g(\imu\om_n, x)|$ must also increase to satisfy Eq.~\eqref{eq:quasi_norm}.
Such correlation is consistent with the behavior of LDOS in Fig.~\ref{fig:mcmillan_bog2}.
On the other hand, in the case of $s$-wave SC with even-frequency pairing, both $g(\imu\om_n, x)$ and $f(\imu\om_n, x)$ are purely imaginary.
Hence, if $|f(\imu\om_n, x)|$ increases, $|g(\imu\om_n, x)|$ decreases.
This tradeoff relation corresponds to the gap in the LDOS in the presence of pair amplitude.
Although the correlation between the LDOS and pair amplitude is demonstrated for bulk, we expect that it holds also for non-uniform cases as supported by the numerical results.

\section{Summary and discussion \label{sec:summary}}

In this paper, we have studied superconductor Josephson junctions with the Bogoliubov Fermi surface, utilizing the low-energy effective model of bogolon.
Since the bogolon Cooper pair arises from the self-energy effect, it is necessary to determine the Green's function satisfying the Gor'kov equation.
For the evaluation of the Green's function, we applied the McMillan's method, with which the non-Hermitian effective Hamiltonian $\hat H_0(x) + \hat \Sigma(\imu\om_n, x)$ needs to be analyzed.
We also calculated the quasiclassical Green's function by eliminating rapidly oscillating components with a scale $k_{\mrm F}^{-1}$ and revealed spatial dependencies.

We compared the results with those of a conventional spin-singlet $s$-wave superconductor and examined the unique characteristics of bogolon Cooper pairs near the interface.
A central difference between the two cases is the frequency dependences of the Cooper pairs.
The bogolons form the odd-frequency pair in bulk and induce the even-frequency pair at the interface.
On the other hand, in the $s$-wave superconductor, the even-frequency pair is realized in bulk and the odd-frequency pair is induced near the interface.
This difference leads to the different $\theta$-dependences of Green functions.
As recapitulated in Table~\ref{tab:comparison}, the bogolon Cooper pair shows distinctive features in physical quantities such as the LDOS and the current-phase relation.

We have studied properties of bogolons (such as bogolons' LDOS, Cooper pairs, and quasiparticle current) in this paper as a first step to understanding the junctions with BFS.
To gain more quantitative insights relevant to existing superconductors, the relation between physical quantities in terms of bogolons and experimental observables needs to be clarified, which remains to be clarified.
A combination of first principles calculations \cite{Miki24} with junctions presents an intriguing avenue for future research \cite{Hirose94, Hirose95}.

\section*{Acknowledgments}

TM is grateful to R. Iwazaki for fruitful discussions.
This work was supported by JSPS with Grants-in-Aid for Scientific research No.~23KJ0298 (TM), No.~23K17668 (YT and SH), No.~24K00583 (YT), No.~21K03459 (SH), and No.~23H01130 (SH).
ST was supported by the W{\"u}rzburg-Dresden Cluster of Excellence ct.qmat, EXC2147, project-id 390858490, the DFG (SFB 1170), and the Bavarian Ministry of Economic Affairs, Regional Development and Energy within the High-Tech Agenda Project ``Bausteine f{\"u}r das Quanten Computing auf Basis topologischer Materialen.''

\appendix

\section{Correspondence between bogolons and original electron degrees of freedom \label{sec:app_origin}}

To see the origin of the pair potential of bogolons, we start with the Hamiltonian of original electrons.
We here consider the impurity effect, which is significant in the low energy region in the presence of the BFS\@~\cite{Miki21}.
For concreteness, we employ the $j = 3/2$ model, which has been discussed in previous studies on the superconductors with BFS \cite{Agterberg17, Brydon18, Tamura20, Oh20, Oh21}.
The impurity potential part of the Hamiltonian is defined in terms of the original electrons:
\begin{align}
  \mathscr H_{\rm imp} &= \sum_{i} \int \diff \bm r \sum_{\eta}
  \mcal U_\eta (\bm r - \bm R_i)
   \vec c^\dg (\bm r) \hat O^{\eta} \vec c (\bm r) \label{eq:himp}
\end{align}
with $\vec c (\bm r) = (c_{3/2}(\bm r), c_{3/2}(\bm r), c_{1/2}(\bm r), c_{-1/2}(\bm r), c_{-3/2}(\bm r))^\T$.
In the above expression, we consider the isotropic ($\eta = 1$) and anisotropic ($\eta =xy,yz,zx,z^2,x^2-y^2$) scattering centers located at $\bm R_i$.

The operators of electrons and those of bogolons are connected by the Bogoliubov transformation, which is given by
\begin{align}
  c_{\bm km} &= u_{\bm km}^* \al_{\bm k} + v_{-\bm k,m} \al_{-\bm k}^\dg \label{eq:bogo_trf},
\end{align}
where $c_{\bm km}$ is a Fourier component of $c_m(\bm r)$ in Eq.~\eqref{eq:himp}.
Using this transformation, we rewrite Eq.~\eqref{eq:himp} as
\begin{align}
  \mathscr H_{\rm imp} &= \frac{1}{V} \sum_{\bm{k} , \bm{q}} \rho_{\bm{q}} U_1(\bm{k} , \bm{q}) \alpha_{\bm{k} + \bm{q}}^\dagger \alpha_{\bm{k}} \notag \\
  &\quad + \frac{1}{V} \sum_{\bm{k} , \bm{q}} \rho_{\bm{q}} U_2(\bm{k} , \bm{q}) \alpha_{\bm{k} + \bm{q}}^\dagger \alpha_{-\bm{k}}^\dagger + \mathrm{H.c.} \notag \\
  &\quad + \mathrm{Const.} \label{eq:himp_bog}
\end{align}
with
\begin{align}
  U_1(\bm{k} , \bm{q}) &= \sum_{\eta} \sum_{m,m'} \mcal U_{\eta}(\bm{q}) [u_{\bm{k} + \bm{q} , m} O^\eta_{mm'} u_{\bm{k} , m'}^\ast \notag \label{eq:u1} \\
  &\qquad - v_{\bm{k} , m}^\ast O_{mm'}^\eta v_{\bm{k} + \bm{q} , m'}], \\
  U_2(\bm{k} , \bm{q}) &= \sum_{\eta} \sum_{m,m'} \mcal U_{\eta}(\bm{q}) u_{\bm{k} + \bm{q} , m} O^\eta_{mm'} v_{-\bm{k} , m'}. \label{eq:u2}
\end{align}
The full list of $4\times 4$ matrices is defined by using the $\hat {\bm J}$ matrix in Ref.~\cite{Tamura20}.
The particle number of bogolons is not conserved because of the presence of $U_2(\bm k, \bm q)$.
We here provide the expression of the pair potential with the Born approximation from Ref~\cite{Miki21}:
\begin{align}
    \Gamma_{2\bm k} = 4\pi D_0 n_{\mrm{imp}} \la U_1(\bm k, \bm q) U_2(\bm k, \bm q) \ra_{\bm q},
\end{align}
where $\la \cdots \ra_{\bm q} = \int\diff\bm q \cdots / \int\diff\bm q 1,\, n_{\mrm{imp}} = V^{-1} \sum_i 1$, and $D_0$ is a DOS at Fermi energy.
Therefore, we need to take into account $U_2(\bm k, \bm q)$ for the presence of pair potential $\Gamma_{2\bm k}$.
Since the phase of the superconductor is given by $\arg uv$, the phase of $\Gamma_{2}$ for bogolons is identical to the phase of the superconductor as discussed in the main text.

We note that a similar discussion is possible for the self-consistent Born approximation \cite{Kopnin_book} by considering that the Green's function is determined self consistently including the anomalous Green's function \cite{Miki21}, and the above conclusion does not change.

\section{McMillan formalism \label{sec:app_mcmillan}}

In this Appendix, we briefly follow the formalism by McMillan \cite{McMillan68, Furusaki91, Tanaka96, Tanaka97, Kashiwaya00, Burset15, Lu18, Tanaka24}.
Firstly, we impose the boundary condition for the Green's function.
For $x < x'$, the boundary condition at $x \to \pm\infty$ is given by
\begin{align}
    \hat G(\imu\om_n, x, x' \to \infty) = \hat G(\imu\om_n, x \to -\infty, x') = 0, \label{eq:boundary_1}
\end{align}
while for $x > x'$, the condition is given by
\begin{align}
    \hat G(\imu\om_n, x \to \infty, x') = \hat G(\imu\om_n, x, x' \to -\infty) = 0. \label{eq:boundary_2}
\end{align}
We chose the wave functions Eq.~\eqref{eq:wfc} and Fig.~\ref{fig:wfc} to satisfy Eqs.~\eqref{eq:boundary_1} and \eqref{eq:boundary_2} \cite{McMillan68, Furusaki91, Tanaka96, Tanaka97, Kashiwaya00, Burset15, Lu18, Tanaka24}.

Although the coefficients $\al_1, \cdots, \al_4, \be_1, \cdots, \be_4$ are uniquely determined by the boundary conditions Eqs.~\eqref{eq:boundary_3} and \eqref{eq:boundary_4}, these are rewritten in a simpler form by using Eqs.~\eqref{eq:uu} and \eqref{eq:vv}.
Consequently, we obtain the following forms \cite{Furusaki91, Tanaka96, Tanaka97, Kashiwaya00, Burset15, Lu18, Tanaka24}:
\begin{align}
    &\al_1 = \frac{m (\imu\om_n + \imu\Gamma_{1\mrm{L}})}{\imu k_{\mrm F} \hbar^2 \Omega_{\mrm{L}}(\imu\om_n)} \frac{c_{\mrm{in}}^{(-)}}{c_{\mrm{in}}^{(+)} c_{\mrm{in}}^{(-)} - d_{\mrm{in}}^{(+)} d_{\mrm{in}}^{(-)}}, \label{eq:al1} \\
    &\al_2 = -\frac{m (\imu\om_n + \imu\Gamma_{1\mrm{L}})}{\imu k_{\mrm F} \hbar^2 \Omega_{\mrm{L}}(\imu\om_n)} \frac{d_{\mrm{in}}^{(-)}}{c_{\mrm{in}}^{(+)} c_{\mrm{in}}^{(-)} - d_{\mrm{in}}^{(+)} d_{\mrm{in}}^{(-)}}, \label{eq:al2} \\
    &\al_3 = -\frac{m (\imu\om_n + \imu\Gamma_{1\mrm{L}})}{\imu k_{\mrm F} \hbar^2 \Omega_{\mrm{L}}(\imu\om_n)} \frac{d_{\mrm{in}}^{(+)}}{c_{\mrm{in}}^{(+)} c_{\mrm{in}}^{(-)} - d_{\mrm{in}}^{(+)} d_{\mrm{in}}^{(-)}}, \label{eq:al3} \\
    &\al_4 = \frac{m (\imu\om_n + \imu\Gamma_{1\mrm{L}})}{\imu k_{\mrm F} \hbar^2 \Omega_{\mrm{L}}(\imu\om_n)} \frac{c_{\mrm{in}}^{(+)}}{c_{\mrm{in}}^{(+)} c_{\mrm{in}}^{(-)} - d_{\mrm{in}}^{(+)} d_{\mrm{in}}^-}, \label{eq:al4}
\end{align}
and
\begin{align}
    &\be_1 = \frac{m (\imu\om_n + \imu\Gamma_{1\mrm{L}})}{\imu k_{\mrm F} \hbar^2 \Omega_{\mrm{L}}(\imu\om_n)} \frac{\tilde c_{\mrm{in}}^{(-)}}{\tilde c_{\mrm{in}}^{(+)} \tilde c_{\mrm{in}}^{(-)} - \tilde d_{\mrm{in}}^{(+)} \tilde d_{\mrm{in}}^{(-)}}, \label{eq:be1} \\
    &\be_2 = -\frac{m (\imu\om_n + \imu\Gamma_{1\mrm{L}})}{\imu k_{\mrm F} \hbar^2 \Omega_{\mrm{L}}(\imu\om_n)} \frac{\tilde d_{\mrm{in}}^{(-)}}{\tilde c_{\mrm{in}}^{(+)} \tilde c_{\mrm{in}}^{(-)} - \tilde d_{\mrm{in}}^{(+)} \tilde d_{\mrm{in}}^{(-)}}, \label{eq:be2} \\
    &\be_3 = -\frac{m (\imu\om_n + \imu\Gamma_{1\mrm{L}})}{\imu k_{\mrm F} \hbar^2 \Omega_{\mrm{L}}(\imu\om_n)} \frac{\tilde d_{\mrm{in}}^{(+)}}{\tilde c_{\mrm{in}}^{(+)} \tilde c_{\mrm{in}}^{(-)} - \tilde d_{\mrm{in}}^{(+)} \tilde d_{\mrm{in}}^{(-)}}, \label{eq:be3} \\
    &\be_4 = \frac{m (\imu\om_n + \imu\Gamma_{1\mrm{L}})}{\imu k_{\mrm F} \hbar^2 \Omega_{\mrm{L}}(\imu\om_n)} \frac{\tilde c_{\mrm{in}}^{(+)}}{\tilde c_{\mrm{in}}^{(+)} \tilde c_{\mrm{in}}^{(-)} - \tilde d_{\mrm{in}}^{(+)} \tilde d_{\mrm{in}}^{(-)}}, \label{eq:be4}
\end{align}
where we have assumed $k_{\mrm L}^\pm \simeq k_{\mrm R}^\pm \simeq k_{\mrm F}$.
Solving Eq.~\eqref{eq:wfc_cond_1} and Eq.~\eqref{eq:wfc_cond_2}, we obtain the coefficients $a_{\mrm{out}}^{(\pm)}$, $a_{\mrm{in}}^{(\pm)}$, $b_{\mrm{out}}^{(\pm)}$, $b_{\mrm{in}}^{(\pm)}$, $c_{\mrm{out}}^{(\pm)}$, $c_{\mrm{in}}^{(\pm)}$, $d_{\mrm{out}}^{(\pm)}$ and $d_{\mrm{in}}^{(\pm)}$, which are given by
\begin{align}
    &a_{\mrm{out}}^{(\pm)} 
    = \frac{\Xi_{\mrm{LR}}^{\pm\pm} \Xi_{\mrm{LR}}^{\pm\mp}}{Z^2 \Xi_{\mrm{LL}}^{\mp\pm} \Xi_{\mrm{RR}}^{\mp\pm} - \Xi_{\mrm{LR}}^{\mp\pm} \Xi_{\mrm{LR}}^{\pm\mp}}, \label{eq:aoutz} \\
    &b_{\mrm{out}}^{(\pm)} 
    = \frac{-Z(Z \pm \imu)\Xi_{\mrm{LL}}^{\mp\pm} \Xi_{\mrm{RR}}^{\mp\pm}}{Z^2 \Xi_{\mrm{LL}}^{\mp\pm} \Xi_{\mrm{RR}}^{\mp\pm} - \Xi_{\mrm{LR}}^{\mp\pm} \Xi_{\mrm{LR}}^{\pm\mp}}, \\
    &c_{\mrm{out}}^{(\pm)} 
    = \frac{(\pm\imu Z - 1)\Xi_{\mrm{LR}}^{\pm\mp} \Xi_{\mrm{LL}}^{\mp\pm}}{Z^2 \Xi_{\mrm{LL}}^{\mp\pm} \Xi_{\mrm{RR}}^{\mp\pm} - \Xi_{\mrm{LR}}^{\mp\pm} \Xi_{\mrm{LR}}^{\pm\mp}}, \\
    &d_{\mrm{out}}^{(\pm)} 
    = \frac{\pm \imu Z \Xi_{\mrm{LR}}^{\pm\pm} \Xi_{\mrm{LL}}^{\mp\pm}}{Z^2 \Xi_{\mrm{LL}}^{\mp\pm} \Xi_{\mrm{RR}}^{\mp\pm} - \Xi_{\mrm{LR}}^{\mp\pm} \Xi_{\mrm{LR}}^{\pm\mp}},
\end{align}
and
\begin{align}
    &a_{\mrm{in}}^{(\pm)} 
    = \frac{\Xi_{\mrm{RL}}^{\pm\pm} \Xi_{\mrm{RL}}^{\pm\mp}}{Z^2 \Xi_{\mrm{LL}}^{\mp\pm} \Xi_{\mrm{RR}}^{\mp\pm} - \Xi_{\mrm{LR}}^{\mp\pm} \Xi_{\mrm{LR}}^{\pm\mp}}, \\
    &b_{\mrm{in}}^{(\pm)} 
    = \frac{-Z(Z \pm \imu)\Xi_{\mrm{RR}}^{\mp\pm} \Xi_{\mrm{LL}}^{\mp\pm}}{Z^2 \Xi_{\mrm{LL}}^{\mp\pm} \Xi_{\mrm{RR}}^{\mp\pm} - \Xi_{\mrm{LR}}^{\mp\pm} \Xi_{\mrm{LR}}^{\pm\mp}}, \\
    &c_{\mrm{in}}^{(\pm)} 
    = \frac{(\pm\imu Z - 1)\Xi_{\mrm{RL}}^{\pm\mp} \Xi_{\mrm{RR}}^{\mp\pm}}{Z^2 \Xi_{\mrm{LL}}^{\mp\pm} \Xi_{\mrm{RR}}^{\mp\pm} - \Xi_{\mrm{LR}}^{\mp\pm} \Xi_{\mrm{LR}}^{\pm\mp}}, \\
    &d_{\mrm{in}}^{(\pm)} 
    = \frac{\pm \imu Z \Xi_{\mrm{RL}}^{\pm\pm} \Xi_{\mrm{RR}}^{\mp\pm}}{Z^2 \Xi_{\mrm{LL}}^{\mp\pm} \Xi_{\mrm{RR}}^{\mp\pm} - \Xi_{\mrm{LR}}^{\mp\pm} \Xi_{\mrm{LR}}^{\pm\mp}} \label{eq:dinz}
\end{align}
with
\begin{align}
    &\Xi_{r r'}^{ss'} = u_r^s v_{r'}^{s'} - u_{r'}^{s'} v_r^s
\end{align}
and $s,s'=\pm,\ r,r'=\mrm{L}, \mrm{R}$.
We note that the coefficients for wave functions with tilde $\tilde a_{\mrm{out}}^{(\pm)}$, $\tilde a_{\mrm{in}}^{(\pm)}$, $\tilde b_{\mrm{out}}^{(\pm)}$, $\tilde b_{\mrm{in}}^{(\pm)}$, $\tilde c_{\mrm{out}}^{(\pm)}$, $\tilde c_{\mrm{in}}^{(\pm)}$, $\tilde d_{\mrm{out}}^{(\pm)}$, and $\tilde d_{\mrm{in}}^{(\pm)}$ can be obtained by substituting $\theta \to -\theta$ in $a_{\mrm{out}}^{(\pm)}$, $a_{\mrm{in}}^{(\pm)}$, $b_{\mrm{out}}^{(\pm)}$, $b_{\mrm{in}}^{(\pm)}$, $c_{\mrm{out}}^{(\pm)}$, $c_{\mrm{in}}^{(\pm)}$, $d_{\mrm{out}}^{(\pm)}$, and $d_{\mrm{in}}^{(\pm)}$, respectively.
For $\Gamma_{1\mrm L} = \Gamma_{1\mrm R}$ and $|\Gamma_{2\mrm L}| = |\Gamma_{2\mrm R}|$, the coefficients $\tilde a_{\mrm{out}}^{(\pm)}$ and $\tilde b_{\mrm{out}}^{(\pm)}$ reduce to Eq.~\eqref{eq:a}.

We here comment on the eigenequation of Eq.~\eqref{eq:eigen_bulk}.
Using Eq.~\eqref{eq:ham_sig_1}--\eqref{eq:biorthogonal}, we obtain the following relations:
\begin{align}
    \tilde u_r^{\pm} u_r^{\pm} = \frac{1}{2} \Bigg(1 \pm \frac{\Omega_r(\imu\om_n)}{\imu\om_n + \imu\Gamma_{1r}\sgn\om_n}\Bigg), \label{eq:uu} \\
    \tilde v_r^{\pm} v_r^{\pm} = \frac{1}{2} \Bigg(1 \mp \frac{\Omega_r(\imu\om_n)}{\imu\om_n + \imu\Gamma_{1r}\sgn\om_n}\Bigg). \label{eq:vv}
\end{align}
We choose the phase of the eigenvector such that $u_r^{\pm}$ is given by
\begin{align}
    u_r^{\pm} = \sqrt{\frac{1}{2} \Bigg(1 \pm \frac{\Omega_r(\imu\om_n)}{\imu\om_n + \imu\Gamma_{1r}}\Bigg)} \label{eq:u}
\end{align}
for the practical calculation.
Then, $v_r^{\pm}$, $\tilde u_r^{\pm}$, and $\tilde v_r^{\pm}$ are uniquely determined under this choice.

\begin{widetext}

\section{Specific expression for $s$-wave superconductor junction \label{sec:app_swave}}

In this Appendix, we summarize the results for electrons in $s$-wave SC as a reference for comparison with the results for the bogolon junction.
The basis is given by $\vec c(x) = (c_{\ua}(x) , c_{\da}^\dg(x))^\T$ where $c_{\sg}(x)$ is an annihilation operator for the spin $\sg = \ua, \da$ electron.
The self-energy is defined by
\begin{align}
    \hat \Sigma(x) = 
    \begin{pmatrix}
        0 & \Delta(x) \\
        \Delta(x)^\ast & 0
    \end{pmatrix} \label{eq:delta_mat}
\end{align}
with
\begin{align}
    &\Delta(x) = 
    \begin{cases}
        \Delta_{\mrm{L}} = |\Delta_{\mrm{L}}| \epn^{\imu\theta_{\mrm L}} \qquad (x < 0), \\
        \Delta_{\mrm{R}} = |\Delta_{\mrm{R}}| \epn^{\imu\theta_{\mrm R}} \qquad (x > 0).
    \end{cases} \label{eq:delta}
\end{align}

Following the McMillan's method discussed in Sec.~\ref{sec:mcmillan} and Appendix~\ref{sec:app_mcmillan}, we obtain the Green's function as follows \cite{McMillan68, Furusaki91, Tanaka96, Tanaka97, Kashiwaya00, Burset15, Lu18, Tanaka24, TextTanaka2021}:
\begin{align}
    \hat G(\imu\om_n, x, x') 
    &= \frac{m}{2\imu k_{\mrm F} \hbar^2 \Omega_{\mrm{L}}(\imu\om_n)}
    \Bigg[ \Big(\epn^{\imu k_{\mrm L}^+ |x - x'|} + \tilde b_{\mrm{out}}^{(+)} \epn^{-\imu k_{\mrm L}^+ (x + x')}\Big)
    \begin{pmatrix}
        \imu\om_n + \Omega_{\mrm L}(\imu\om_n) & \Delta_{\mrm L} \\
        \Delta_{\mrm L}^\ast & \imu\om_n - \Omega_{\mrm L}(\imu\om_n)
    \end{pmatrix} \nt
    &+ \tilde a_{\mrm{out}}^{(+)} \epn^{-\imu k_{\mrm L}^+ x + \imu k_{\mrm L}^- x'}
    \begin{pmatrix}
        |\Delta_{\mrm L}| &  \epn^{\imu\theta_{\mrm L}} [\imu\om_n + \Omega_{\mrm L}(\imu\om_n)] \\
        \epn^{-\imu\theta_{\mrm L}} [\imu\om_n - \Omega_{\mrm L}(\imu\om_n)] & |\Delta_{\mrm L}|
    \end{pmatrix} \nt
    &+ \Big(\epn^{-\imu k_{\mrm L}^- |x - x'|} + \tilde b_{\mrm{out}}^{(-)} \epn^{\imu k_{\mrm L}^- (x + x')} \Big)
    \begin{pmatrix}
        \imu\om_n - \Omega_{\mrm L}(\imu\om_n) & \Delta_{\mrm L} \\
        \Delta_{\mrm L}^\ast & \imu\om_n + \Omega_{\mrm L}(\imu\om_n)
    \end{pmatrix} \nt
    &+ \tilde a_{\mrm{out}}^{(-)} \epn^{\imu k_{\mrm L}^- x - \imu k_{\mrm L}^+ x'}
    \begin{pmatrix}
        |\Delta_{\mrm L}| &  \epn^{\imu\theta_{\mrm L}} [\imu\om_n - \Omega_{\mrm L}(\imu\om_n)] \\
        \epn^{-\imu\theta_{\mrm L}} [\imu\om_n + \Omega_{\mrm L}(\imu\om_n)] & |\Delta_{\mrm L}|
    \end{pmatrix}
    \Bigg] \label{eq:mcmillan_s}
\end{align}
with
\begin{align}
    &\tilde a_{\mrm{out}}^{(\pm)}(\imu\om_n) 
    = -\frac{|\Delta_{\mrm L}| [\imu\om_n \sin^2(\theta/2) \pm \imu\Omega_{\mrm L}(\imu\om_n) \sin(\theta/2)\cos(\theta/2)]}{Z^2\Omega_{\mrm L}(\imu\om_n)^2 + (\imu\om_n)^2 - |\Delta_{\mrm L}|^2\cos^2(\theta/2)}, \label{eq:a_s} \\
    &\tilde b_{\mrm{out}}^{(\pm)}(\imu\om_n) 
    = -\frac{Z(Z \pm \imu\sgn\om_n) \Omega_{\mrm L}(\imu\om_n)^2 }{Z^2 \Omega_{\mrm L}(\imu\om_n)^2 + (\imu\om_n)^2 - |\Delta_{\mrm L}|^2\cos^2(\theta/2)}. \label{eq:b_s}
\end{align}

We turn to the discussion of the quasiclassical representation, which corresponds to Sec.~\ref{sec:quasiclassical} for the bogolon junction.
The quasiclassical Green's function is given by \cite{Nagato93}
\begin{align}
    \hat g^{\pm\pm}(\imu\om_n, x)
    &= \frac{\imu}{\Omega_{\mrm L}(\imu\om_n)} 
    \Big[\imu\Delta_{\mrm L} \hat \tau^y + \imu\om_n \hat \tau^z 
    + \tilde a_{\mrm{out}}^{(\mp)} \epn^{-\imu k_{\mrm F}(\Omega_{\mrm L}(\imu\om_n)/\mu) x} \Big( \mp\Omega_{\mrm L}(\imu\om_n) \hat \tau^x
    + \imu\om_n \imu \hat \tau^y 
    + |\Delta_{\mrm L}| \hat \tau^z \Big) \Big]. \label{eq:quasiclassical_s}
\end{align}
where we set $\theta_{\mrm L} = 0$ as done for the bogolon junction in Sec.~\ref{sec:quasiclassical}.
The spatial dependence is determined by the factor
\begin{align}
    \exp(2\sqrt{\om_n^2 + |\Delta_{\mrm L}|^2} x / \hbar v_{\mrm F}). \label{eq:exp_s}
\end{align}
In the retarded Green's function, the factor is given by
\begin{align}
    \exp(-2\imu\sqrt{\om^2 - |\Delta_{\mrm L}|^2}\sgn\om x / \hbar v_{\mrm F}). \label{eq:exp_s_ret}
\end{align}
Using Eq.~\eqref{eq:quasiclassical_s}, we obtain the pair amplitudes $f_s(\imu\om_n, x)$ and $f_p(\imu\om_n, x)$ defined by Eqs.~\eqref{eq:quasiclassical_fs_def} and \eqref{eq:quasiclassical_fp_def}, which are given by
\begin{align}
    f_s(\imu\om_n, x) 
    &= \frac{2\imu\Delta_{\mrm L}}{\Omega_{\mrm L}(\imu\om_n)}
    +\bigg[\imu\Big(\tilde a_{\mrm{out}}^{(+)} - \tilde a_{\mrm{out}}^{(-)}\Big)
    + \imu\Big(\tilde a_{\mrm{out}}^{(+)} + \tilde a_{\mrm{out}}^{(-)}\Big)
    \frac{\imu\om_n}{\Omega_{\mrm L}(\imu\om_n)} \bigg]
    \epn^{-\imu k_{\mrm F}(\Omega_{\mrm L}(\imu\om_n)/\mu) x},
    \label{eq:quasiclassical_fs_s} \\
    f_p(\imu\om_n, x) 
    &= \bigg[
    \imu\Big(\tilde a_{\mrm{out}}^{(+)} + \tilde a_{\mrm{out}}^{(-)}\Big)
    - \imu\Big(\tilde a_{\mrm{out}}^{(+)} - \tilde a_{\mrm{out}}^{(-)}\Big)
    \frac{\imu\om_n}{\Omega_{\mrm L}(\imu\om_n)}
    \bigg] \epn^{-\imu k_{\mrm F}(\Omega_{\mrm L}(\imu\om_n)/\mu) x}. 
    \label{eq:quasiclassical_fp_s}
\end{align}
We also calculate the LDOS and $j(\imu\om_n, x)$ in quasiclassical representation.
The LDOS is expressed as
\begin{align}
    D(\om, x) 
    &= -\frac{1}{\pi}\real\bigg[\frac{2(\om + \imu0^+)}{\Omega_{\mrm{ret}}(\omega)}
    + \frac{\tilde a_{\mrm{out, ret}}^{(+)} + \tilde a_{\mrm{out, ret}}^{(-)}}{\Omega_{\mrm{ret}}(\omega)}
    |\Delta_{\mrm L}| \epn^{-\imu k_{\mrm F}(\Omega_{\mrm{ret}}(\omega)/\mu) x}\bigg],
    \label{eq:quasiclassical_ldos_s}
\end{align}
where $a_{\mrm{out, ret}}^{(\pm)}$ is a retarded version of $a_{\mrm{out}}^{(\pm)}$.
$j(\imu\om_n, x)$ defined by Eq.~\eqref{eq:quasiclassical_j_def} is given by
\begin{align}
    j(\imu\om_n, x) 
    &= \frac{|\Delta_{\mrm L}|}{\Omega_{\mrm L}(\imu\om_n)}
    \Big(\tilde a_{\mrm{out}}^{(+)} - \tilde a_{\mrm{out}}^{(-)}\Big) 
    \epn^{-\imu k_{\mrm F}(\Omega_{\mrm L}(\imu\om_n)/\mu) x}. 
    \label{eq:quasiclassical_j_s}
\end{align}

\section{Quasiclassical Green's function for bogolon junction \label{sec:app_quasiclassical}}

In this Appendix, we list the specific form of the physical quantities in the quasiclassical representations.
The physical quantities discussed in Sec.~\ref{sec:result} are expressed as the combination of $g^{++}$ and $g^{--}$ \cite{Nagato93}. 
The $s$-wave and $p$-wave components of the pair amplitudes are given by
\begin{align}
    &f_s(\imu\om_n, x) 
    = [\hat g^{++}(\imu\om_n, x) + \hat g^{--}(\imu\om_n, x)]_{12}, \label{eq:quasiclassical_fs_def} \\
    &f_p(\imu\om_n, x) = [\hat g^{++}(\imu\om_n, x) - \hat g^{--}(\imu\om_n, x)]_{12}. \label{eq:quasiclassical_fp_def}
\end{align}
The LDOS of bogolons and $j(\imu\om_n, x)$ in the quasiclassical representations are expressed as
\begin{align}
    &D(\om, x) 
    = -\frac{1}{\pi}\imag[\hat g^{++}(\om + \imu 0^+, x) + \hat g^{--}(\om + \imu 0^+, x)]_{11}, \label{eq:quasiclassical_ldos_def} \\
    &j(\imu\om_n, x) = \frac{\imu}{n_{\mrm d}} [\hat g^{++}(\imu\om_n, x) - \hat g^{--}(\imu\om_n, x)]_{11}, \label{eq:quasiclassical_j_def}
\end{align}
respectively.

Below, we list the expressions of physical quantities derived from the quasiclassical Green's function Eq.~\eqref{eq:quasi1}. 
The $s$-wave and $p$-wave components of the pair amplitudes Eqs.~\eqref{eq:quasiclassical_fs_def} and \eqref{eq:quasiclassical_fp_def} are reduced to
\begin{align}
    f_s(\imu\om_n, x) &= \frac{-2\imu\Gamma_2\sgn\om_n}{\Omega_{\mrm L}(\imu\om_n)}
    +\bigg[-\Big(\tilde a_{\mrm{out}}^{(+)} - \tilde a_{\mrm{out}}^{(-)}\Big)
    - (\tilde a_{\mrm{out}}^{(+)} + \tilde a_{\mrm{out}}^{(-)})
    \frac{\imu\om_n + \imu\Gamma_{1\mrm L}\sgn\om_n}{\Omega_{\mrm L}(\imu\om_n)} \bigg]
    \epn^{-\imu k_{\mrm F}(\Omega_{\mrm L}(\imu\om_n)/\mu) x},
    \label{eq:quasiclassical_fs}
\end{align}
and
\begin{align}
    f_p(\imu\om_n, x)
    &= \bigg[
    \Big(\tilde a_{\mrm{out}}^{(+)} + \tilde a_{\mrm{out}}^{(-)}\Big)
    + \Big(\tilde a_{\mrm{out}}^{(+)} - \tilde a_{\mrm{out}}^{(-)}\Big)
    \frac{\imu\om_n + \imu\Gamma_{1\mrm L}\sgn\om_n}{\Omega_{\mrm L}(\imu\om_n)}
    \bigg] \epn^{-\imu k_{\mrm F}(\Omega_{\mrm L}(\imu\om_n)/\mu) x},
    \label{eq:quasiclassical_fp}
\end{align}
respectively.
We note that $\tilde a_{\mrm{out}}^{(+)} + \tilde a_{\mrm{out}}^{(-)}$ is an even-function with respect to frequency, while $\tilde a_{\mrm{out}}^{(+)} - \tilde a_{\mrm{out}}^{(-)}$ is an odd-function.
Furthermore, comparing Eq.~\eqref{eq:quasiclassical_fs} with Eq.~\eqref{eq:quasiclassical_fp}, the positions of two factors $\tilde a_{\mrm{out}}^{(+)} + \tilde a_{\mrm{out}}^{(-)}$ and $\tilde a_{\mrm{out}}^{(+)} - \tilde a_{\mrm{out}}^{(-)}$ are reversed.
Therefore, we can check that Eq.~\eqref{eq:quasiclassical_fs} corresponds to the odd-frequency pair amplitude, while Eq.~\eqref{eq:quasiclassical_fp} corresponds to the even-frequency pair amplitude even in the quasiclassical representation.

The LDOS of bogolons is given by
\begin{align}
    D(\om, x) 
    &= -\frac{1}{\pi}\real\bigg[\frac{2(\om + \imu\Gamma_{1\mrm L})}{\Omega_{\mrm{ret}}(\omega)}
    + \frac{\tilde a_{\mrm{out, ret}}^{(+)} + \tilde a_{\mrm{out, ret}}^{(-)}}{\Omega_{\mrm{ret}}(\omega)}
    \imu|\Gamma_{2\mrm L}| \epn^{-\imu k_{\mrm F}(\Omega_{\mrm{ret}}(\omega)/\mu) x}\bigg],
    \label{eq:quasiclassical_ldos}
\end{align}
where $a_{\mrm{out, ret}}^{(\pm)}$ is a retarded version of $a_{\mrm{out}}^{(\pm)}$.
$j(\imu\om_n, x)$ is expressed as
\begin{align}
    j(\imu\om_n, x) 
    = -\frac{|\Gamma_{2\mrm L}|\sgn\om_n}{2\imu\Omega_{\mrm L}(\imu\om_n)}
    \Big(\tilde a_{\mrm{out}}^{(+)} - \tilde a_{\mrm{out}}^{(-)}\Big)
    \epn^{-\imu k_{\mrm F}(\Omega_{\mrm L}(\imu\om_n)/\mu) x}.
    \label{eq:quasiclassical_j}
\end{align}

\end{widetext}

\bibliography{bogolon_junction}
\bibliographystyle{apsrev4-2}

\end{document}